\documentclass[traditabstract]{aa}
\usepackage{longtable}

\usepackage{graphicx}
\usepackage{txfonts}
\usepackage{natbib}
\bibpunct{(}{)}{;}{a}{}{,} % to follow the A&A style

\def\solar {\ifmmode_{\mathord\odot} \else $_{\mathord\odot}$ \fi}% _solar
\def\jup {\ifmmode_{\mathrm{Jup}} \else $_{\mathrm{Jup}}$ \fi}% _jup
\def\earth {\ifmmode_{\mathord\oplus} \else $_{\mathord\oplus}$ \fi}% _earth
\def\Msol {\ifmmode {\,\mathrm{M}\solar} \else \,M\solar \fi}     % solar mass
\def\Rsol {\ifmmode {\,\mathrm{R}\solar} \else R\solar \fi}     % solar radius
\def\Lsol {\ifmmode {\,\mathrm{L}\solar} \else L\solar \fi}     % solar radius
\def\Mjup {\ifmmode {\,\mathrm{M}\jup} \else M\jup \fi}
\def\Mearth {\ifmmode {\,\mathrm{M}\earth} \else M\earth \fi}

\def\mps {\ifmmode {\,\mathrm{m\,s^{-1}}} \else $\mathrm{m\,s^{-1}}$ \fi}     % meter per sec

\def\k{p}
\def\l{k}
\def\SS{\left({\cal S}\right)}

\begin{document}
  \title{The HARPS search for southern extra-solar planets\thanks{Based on observations made with the HARPS instrument on the ESO 3.6 m telescope under the program IDs 072.C-0488, 082.C-0718 and 183.C-0437 at Cerro La Silla (Chile). Radial-velocity time series are available in electronic format the CDS 
via anonymous ftp to cdsarc.u-strasbg.fr (130.79.128.5) or via 
http://cdsweb.u-strabg.fr/cgi-bin/qcat?J/A+A/
}}

  \subtitle{XXXIV. A planetary system around the nearby M dwarf \object{GJ~163}, with a super-Earth possibly in the habitable zone.}

\author{X.~Bonfils   \inst{1} 
   \and G.~Lo~Curto     \inst{2}
   \and A.~C.~M.~Correia \inst{3,11}
 \and J.~Laskar \inst{11}
  \and S.~Udry \inst{5}
 \and X.~Delfosse \inst{1}
 \and T.~Forveille \inst{1}
 \and N.~Astudillo-Defru \inst{1}
 \and W.~Benz \inst{6}
 \and F.~Bouchy \inst{5,7,8,9}
 \and M.~Gillon \inst{10}
 \and G.~H\'ebrard \inst{7,9}
 \and C.~Lovis  \inst{5}
  \and M.~Mayor \inst{5}
 \and C.~Moutou \inst{8}
 \and D.~Naef \inst{5}
 \and V.~Neves \inst{1,4,12}
 \and F.~Pepe \inst{5}
 \and C.~Perrier \inst{1}
 \and D.~Queloz \inst{5}
 \and N.~C.~Santos \inst{4,12}
 \and D.~S\'egransan \inst{5}
}

  \offprints{X. Bonfils}
	\institute{UJF-Grenoble 1 / CNRS-INSU, Institut de Plan\'etologie et d'Astrophysique de Grenoble (IPAG) UMR 5274, Grenoble, F-38041, France
           \and European Southern Observatory, Karl-Schwarzschild-Str. 2, D-85748 Garching bei M\"unchen, Germany            
           \and Department of Physics, I3N, University of Aveiro, Campus Universit\'ario de Santiago, 3810-193 Aveiro, Portugal
	  \and Centro de Astrof\'{i}sica, Universidade do Porto, Rua das Estrelas, 4150-762 Porto, Portugal
           \and Observatoire de Gen\`eve, Universit\'e de Gen\`eve, 51 ch. des Maillettes, 1290 Sauverny, Switzerland
           \and Physikalisches Institut, Universitat Bern, Silderstrasse 5, CH-3012 Bern, Switzerland
	  \and Institut d'Astrophysique de Paris, CNRS, Universit\'e Pierre et Marie Curie, 98bis Bd Arago, 75014 Paris, France
           \and Laboratoire d'Astrophysique de Marseille, UMR 6110 CNRS, UniversitŽ de Provence, 38 rue Fr\'ed\'eric Joliot-Curie, 13388 Marseille Cedex 13, France
           \and Observatoire de Haute-Provence, 04870 Saint-Michel l'Observatoire, France
	  \and Institut dÕAstrophysique et de G\'eophysique, Universit\'e de Li\`ege, All\'ee du 6 Ao\^{u}t 17, Bat. B5C, 4000 Li\`ege, Belgium
           \and Astronomie et Syst\`emes Dynamiques, IMCCE-CNRS UMR8028, Observatoire de Paris, UPMC, 77 Av. Denfert-Rochereau, 75014~Paris, France
	  \and Departamento de F\'{i}sica e Astronomia, Faculdade de Ci\^encias, Universidade do Porto, Rua do Campo Alegre, 4169-007 Porto, Portugal
           }

  \date{Received August 15, 2012 / Accepted June 4th, 2013}

 \abstract
{ The meter-per-second precision achieved by today velocimeters enables the search for $1-10~\rm{M_\oplus}$ planets in the habitable zone of cool stars.
This paper reports on the detection of 3 planets orbiting \object{GJ 163} (HIP19394), a M3 dwarf monitored by our ESO/HARPS search for planets.
We made use of the HARPS spectrograph to collect 150 radial velocities of GJ 163 over a period of 8 years. 

We searched the RV time series for coherent signals and found 5 distinct periodic variabilities. We investigated the stellar activity and casted doubts on the planetary interpretation for 2 signals. 
Before more data can be acquired we concluded that at least 3 planets are orbiting GJ\,163.
They have orbital periods of $P_b=8.632\pm0.002$, $P_c=25.63\pm0.03$ and $P_d=604\pm8$ days and minimum masses $m \ sin i = 10.6\pm0.6$, $6.8\pm0.9$, and $29\pm3$ M$_\oplus$, respectively. We hold our interpretations for the 2 additional signals with periods $P_{(e)}=19.4$ and $P_{(f)}=108$ days.

The inner pair presents an orbital period ratio of $2.97$, but a dynamical analysis of the system shows that it lays outside the 3:1 mean motion resonance.
GJ\,163c, in particular, is a super-Earth with an equilibrium temperature of $T_{\rm eq} = (302\pm10) (1-A)^{1/4}$~K and may lie in the so called habitable zone for albedo values ($A=0.34-0.89$) moderately higher than that of Earth ($A_\oplus=0.2-0.3$). 
}

\titlerunning{Low-mass planets around the nearby M dwarf \object{GJ~163}}
\authorrunning{X. Bonfils et al.}

  \maketitle

 %________________________________________________________________

\section{Introduction}
In 15 years or so, we have witnessed impressive progresses in radial-velocity measurements. One spectrograph in particular -- the High Accuracy Radial velocity Planet Searcher \citep[HARPS; ][]{Mayor:2003} -- broke the former 3-m/s precision floor and enabled the detection of exoplanets in a yet unknown mass-period domain.

Notably, planets with masses below 10 M$_{\oplus}$ and equilibrium temperatures possibly between $\sim175-270$ K (for plausible albedos) have started to be detected. That subset of detections includes GJ\,581d \citep{Udry:2007a, Mayor:2009}, HD\,85512b \citep{Pepe:2011b} and GJ\,667Cc \citep{Bonfils:2011b, Delfosse:2012} which lie in the so called Habitable Zone (HZ) of their host star. Depending on the nature of their atmospheres, liquid water may flow on their surface and, because liquid water is thought as a prerequisite for the emergence of life as we know it, these planets constitute a prized sample for further characterization of their atmosphere and the search for possible biosignatures.

The present paper reports on the detection of at least 3 planets orbiting the nearby M dwarfs GJ\,163. One of them, GJ\,163c, might be of particular interest in term of habitability. Our report is structured as follows. Sect.~\ref{sect:prop} profiles the host star GJ\,163. Sect~\ref{sect:obs} briefly describes the collection of radial-velocity data. Sect~\ref{sect:data} presents our orbital analysis based on both a Markov Chain Monte Carlo and periodogram algorithms. Then, we investigate more closely which signal could result from stellar activity rather than planets (Sect.~\ref{sect:activity}) and retain a solution with 3 planets.
We next investigate the role of planet-planet interactions in the system (Sect.~\ref{sect:dyn}) and in particular whether planets b and c participate in a resonance. Sect~\ref{sect:hab} discusses GJ\,163c in term of habitability before we present our conclusions in Sect.~\ref{sect:concl}.

\section{\label{sect:prop}The properties of \object{GJ~163}}

\object{GJ~163} (\object{HIP 19394}, \object{LHS 188}) is a M3.5 dwarf \citep{Hawley:1996}, distant of  $15.0\pm0.4~\mathrm{pc}$ \citep[$\pi = 66.69 \pm 1.82~\mathrm{mas}$ --][]{Leeuwen:2007} and seen in the Doradus constellation ($\alpha=04^h09^m16^s$, $\delta=-53^\circ22^\prime23^{\prime\prime}$).

Its photometry \citep[$V =  11.811 \pm 0.012$; $K = 7.135 \pm 0.021$ --][]{Koen:2010, Cutri:2003} and parallax imply absolute magnitudes of $M_V = 10.93 \pm 0.14$ and $M_K = 6.26 \pm 0.14$. \object{GJ 163}'s $J-K$ color \citep[$=0.813$ --][]{Cutri:2003} and the \citet{Leggett:2001} color-bolometric relation result in a K-band bolometric correction of $BC_K=2.59\pm0.07$, and in a $L_\star=0.022\pm0.002$ \Lsol luminosity, in good agreement with \citet{Casagrande:2008}'s direct determination ($M_{bol}=8.956$; $L_\star=0.021$). The K-band mass-luminosity relation of \cite{Delfosse:2000} gives a $0.40 \Msol$ mass with a $\sim10\%$ uncertainty. 

Its UVW galactic velocities place \object{GJ\,163} between the old disk and halo populations \citep{Leggett:1992}. We refined GJ\,163's UVW velocities using both the systemic velocity we measured from HARPS spectra (Table \ref{table:fit}) and proper motion from Hipparcos \citep{Leeuwen:2007}. We obtained U=69.7, V=$-$76.0 and, W=1.2 km/s, that confirmed a membership to an old dynamical population. 

Stellar metallicity is known to be statistically related to dynamical populations. For the Halo population, the metallicity peaks at [Fe/H]$\sim-$1.5 \citep{Ryan:1991} whereas that of the Old Disk peaks at [Fe/H]$\sim-$0.7 \citep{Gilmore:1995}. The widths of these distributions are however wide and both populations have a small fraction of stars with solar metallicity. \citet{Casagrande:2008} attributes a metallicity close to that of the median of the solar neighborhood to GJ\,163 ([Fe/H]=$-$0.08). And \citet{Schlaufman:2010}'s photometric relation \citep[or its slight update by][]{Neves:2012} finds a quasi-solar metallicity of [Fe/H] = $-$0.01. It is therefore difficult to conclude whether GJ\,163 belongs to the metal-rich tail of an old population or is a younger star accelerated
to the typical galactic velocity of an old population.

\begin{figure}
\centering
\includegraphics[width=1.1\linewidth]{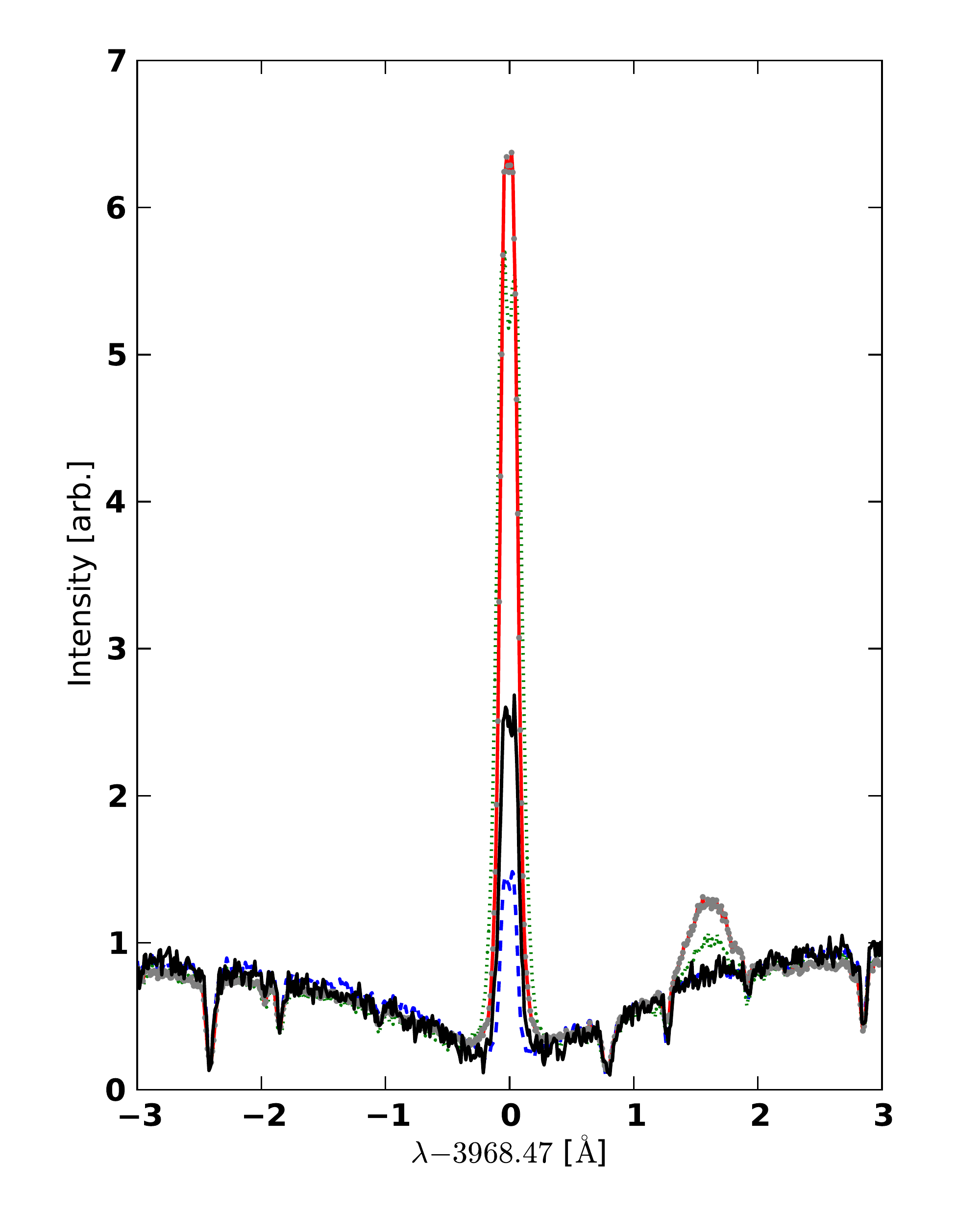}
       \caption{\label{fig:CaIIH}Emission reversal in the 
       \ion{Ca}{ii} H line of \object{GJ~674} (red line; M2.5V; $P_{rot}$=35~d), \object{GJ\,176} (green dots; M2V; $P_{rot}$=39~d), \object{GJ\,163} (black line; M3.5) and,
       \object{GJ 581} (blue dashes; M3V; $P_{rot}$=94~d), ordered from the most prominent to the least prominent peaks. GJ\,163 displays a rather low activity level, which is a strong indication for a slow rotation.       }
       \label{fig:caii}
\end{figure}

GJ\,163 is not detected in the ROSAT All-Sky Survey. We thus used the survey sensitivity limit \citep[$2.4 \times 10^{25} d_{pc}^2$ erg/s --][]{Schmitt:1995} to estimate $\log L_X < 5.39 \times 10^{27}$ erg/s which, given GJ\,163's bolometric luminosity, translates to $R_X = \log L_X/L_{BOL} < -4.17$. For an M dwarf of $\sim0.4$ M$_{\odot}$ the $R_X$ versus
rotation period of \citet{Kiraga:2007} give $P_{rot}>40$ days for such level of X flux. To obtain a better estimate of the rotation period we compared Ca H \& K chromospheric emission lines of GJ~163 with those of 3 other M-dwarf planet hosts with comparable spectral types and known rotational periods : GJ~176 \citep[M2V; $P_{rot}$=39~d --][]{Forveille:2009}, GJ~674 \citep[M2.5V; $P_{rot}$=35~d --][]{Bonfils:2007b} and GJ~581 \citep[M3V; $P_{rot}$=94~d --][]{Vogt:2010}. In figure \ref{fig:CaIIH} we show Ca emission for each stars. GJ~163 has an activity level close to that of GJ~581 which is a very quiet M dwarfs. GJ~163 is much quieter than the 35-40 days rotational period M dwarfs (GJ~176 and GJ~674) and should have rotational period close to that of GJ~581.

\begin{table}
\centering
\caption{
\label{table:stellar}
Observed and inferred stellar parameters for GJ~163}
\begin{tabular}{l@{}lc}
\hline
Spectral Type   &                & M3.5\\
V                       &               & $11.811 \pm 0.012$ \\
$\pi$           &[mas]          & $66.69 \pm 1.82$ \\
Distance                &[pc]           & $15.0 \pm 0.4  $\\
$M_V$           &               & $10.93 \pm 0.06$ \\
K                       &               & $7.135 \pm 0.021$\\
$M_K$           &               & $6.26 \pm 0.06 $\\
$L_\star$       & [$\mathrm{L_\odot}$]          &  $0.022\pm0.003$\\
$dv_r/dt$              & [m\,s$^{-1}$yr$^{-1}$] & $0.491\pm0.013$\\
$M_\star$       & [$\Msol$]             & $ 0.40 \pm 0.04$\\
age & [Gyr] & 1-10  \\
\hline
\end{tabular}
\end{table}

\section{\label{sect:obs}Observations}
We observed \object{GJ 163} with HARPS, a spectrograph fiber-fed by the ESO/3.6m telescope of La Silla Observatory \citep{Mayor:2003, Pepe:2004}. Our settings and computation of radial velocities (RV) remained the same as for our GTO program and we refer the reader to \citet{Bonfils:2011b} for a detailed description. We gathered RVs for 154 epochs spread over 2988 days (8.2 years) between UT 30 October 2003 and 04 January 2012. Table~\ref{tab:rv} (available in electronic form) lists all RVs in the barycentric reference frame of the Solar System. Four measurements have significantly higher uncertainties (the RVs taken at epochs BJD=2454804.7,  2455056.9,  2455057.9 and,  2455136.8 have uncertainties greater than twice the median uncertainty). We removed them and perform our analysis with the remaining 150 RVs.

The proper motion of GJ 163 ($\mu = 1.194\pm0.002$ arcsec/yr) implies a secular change in the orientation of its velocity vector. This results in an apparent radial acceleration $dv/dt = 0.491\pm0.013$~m/s/yr \citep[e.g. ][]{Kurster:2003}, that we subtracted to the RVs listed in Table~\ref{tab:rv} prior to our analysis.

\begin{figure}
\includegraphics[width=\linewidth]{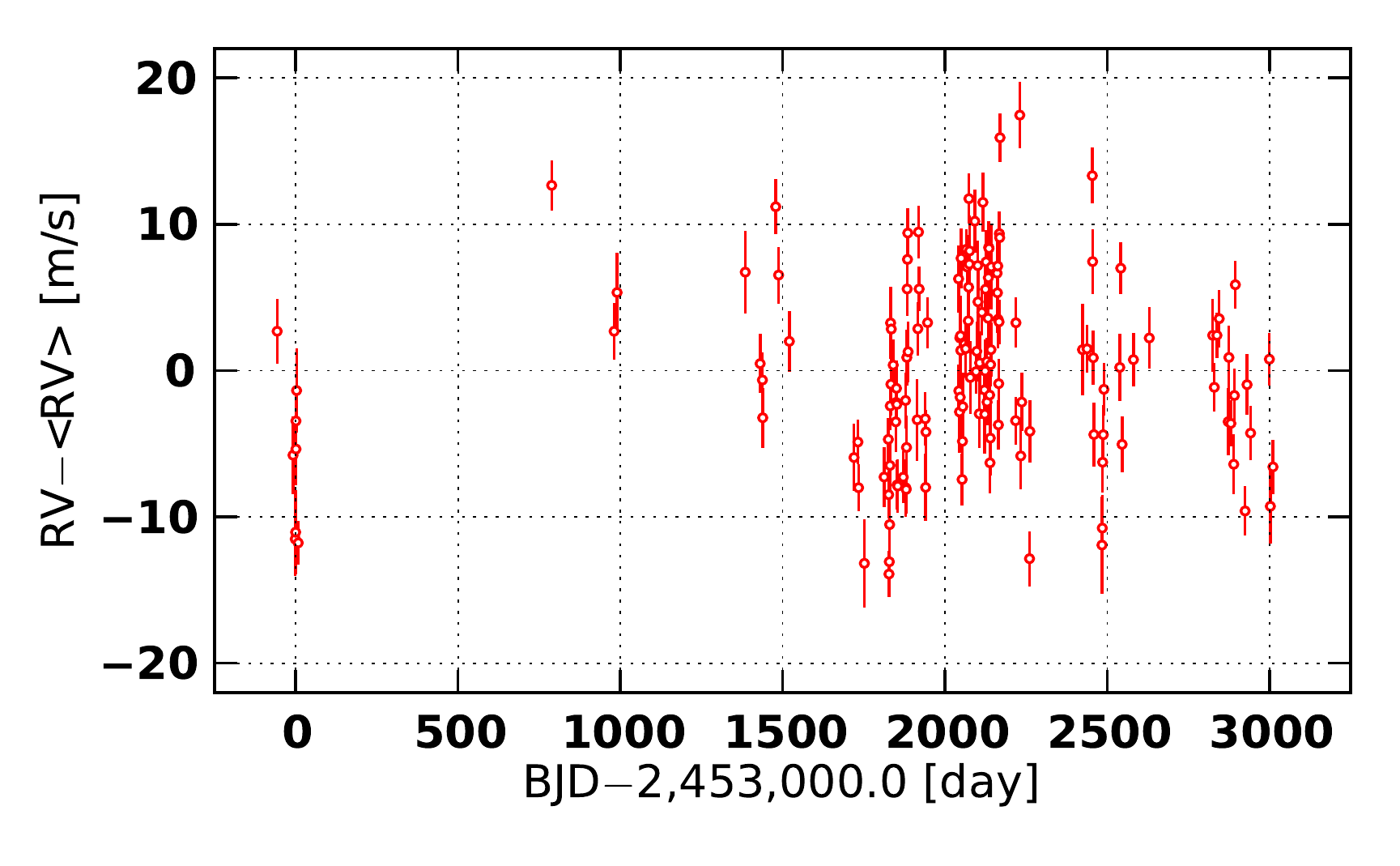}
\caption{RV time series of GJ\,163.}
\label{fig1}
\end{figure}

\section{\label{sect:data}Radial-velocity analysis}

The RV variability of GJ163 ($\sigma_e=6.31$ m/s) is unexplained by photon noise and instrumental errors combined, which are expected to account for a $\sigma_i\sim2.8~{\rm m/s}$ dispersion only \citep[see Sect. 3 in ][]{Bonfils:2011b}. We therefore analyzed the time series and found that this excess of variability results from up to 5 different superimposed signals. We describe below our analysis, made in a Bayesian framework using a Markov Chain Monte Carlo algorithm (\S~\ref{subsect:mcmc}). We also report that similar results are obtained with a classical periodogram analysis (\S~\ref{subsect:periodo}).

\subsection{\label{subsect:mcmc} MCMC modeling}

We used a Markov Chain Monte Carlo algorithm \citep[MCMC][]{Gregory:2005, Gregory:2007, Ford:2005} to sample the joint probability distribution of the model parameters. MCMCs start with random values for all free parameters of a model. Then this solution evolves at the manner of a random walk : each iteration attempts to change the solution randomly, subsequent iterations are accepted following a pseudo-random process and all accepted solutions form the so called {\it chain} of solutions. 

More precisely, for each iteration, we generated a new solution and computed its posterior probability. The posterior probability is the product of the likelihood (the probability of observing the data given the parameter values) by the prior probability of the parameter values. The new solution was accepted with a probability that is function of the ratio between its posterior probability and the posterior probability of the previous solution, such that solutions with a higher posteriori probability were accepted more often. Step-by-step, accepted solutions built a chain which, after enough iterations, reached a stationary state. We then discarded the first iterations and kept only the stationary part of the chain. The distributions of parameter values of all the remaining chain links then corresponded to the targeted joint probability distribution for the model parameters. 

Our implementation closely follows that of Gregory (2007) with several (10 in our case) chains running in parallel. Each chain was attributed a parameter $\beta$ that scaled the likelihood such that chains with a lower $\beta$ value presented a higher acceptance probability. We also paused the MCMC iteration after every 10 steps and proposed the chains to permute their solutions (which was again accepted pseudo-randomly and according to the posterior likelihood ratio between solutions). This approach is reminiscent of simulated annealing algorithms and permits evasion outside of local minima and better exploration of the wide parameter space. Only the chain with $\beta$ = 1 corresponds to the targeted probability distribution. Eventually, we thus discarded all chains but the one with $\beta$ = 1. We adopted the median of the posterior distributions for the optimal parameter values, and the 68\% centered interval for their uncertainties.

We fitted the data with different models. We chose a model without planet where the sole free parameter is the systemic velocity. We also chose models composed of either one, two, three, four, five and six planets on Keplerian orbits. We ran our MCMC algorithm to build chains of 500'000 links and eventually removed the first 10'000 iterations. 

Table~\ref{table:fit} reports optimal parameter values and uncertainties for the model composed of 3 planets. The parameter values are the median of the posterior distributions and the uncertainties are the 68.3\% centered intervals (equivalent to 1-$\sigma$ for gaussian distributions). Notably, the orbital periods of the 3 planets are $P_b=8.631\pm0.002$, $P_c=25.63\pm0.03$ and $P_d=604\pm8$ days. Further assuming a mass $M_\star=$0.4 M$_\odot$ for the primary we estimated their minimum masses to $m \ sin i = 10.6\pm0.6$, $6.8\pm0.9$ and, $29\pm3$ M$_\oplus$, respectively\footnote{An additional $\sim$10\% uncertainties should be added quadratically to the mass uncertainty when accounting for the $\sim$10\% stellar-mass uncertainty.}. When we fitted the data with a model composed of only one planet we found 'b' and when we did with a model composed of two planets we found both planets 'b' and 'd'. When we tried a more complex model composed of 4 or 5 planets, we recovered the Keplerian orbits described in the 3-planet model as well as Keplerian orbits with periods $P_{(e)}=19.4$ and $P_{(f)}=108$ days. And for the most complex model, with 6 planets, the parameters never converged to a unique solution. The 6th orbit is found with orbital periods around 37, 42, 75, 85 and 134 days and, for few thousands chain links, the 19.4-day period is not part of the solution but replaced by one of the orbital periods found for the 6th planet.

More complex models include more free parameters and thus always lead to better fits (i.e. to higher likelihood). To choose whether the improvement in modeling the data justify the additional complexity, we computed Bayes ratios between the different models. They lead to the posterior probability of 1-, 2- and 3-planet models over none-, 1- and 2-planet models to be as high as 10$^{16}$, 10$^{11}$ and 10$^{7}$, respectively, whereas the posterior probabilities for the models with 4, 5 and 6 planets over the models with 3, 4 and 5 planets were only 75, 62 and 5, respectively.  We require that more complex models have a Bayes ratio $>$100 to be accepted and thus concluded that our data show strong evidence for at least 3 planetary signals, and perhaps some evidence for more planets. 

\subsection{\label{subsect:periodo}Periodogram analysis}

We now present an alternative analysis of the radial-velocity time series based on periodograms. We used floating-mean Lomb-Scargle periodograms \citep{Lomb:1976, Scargle:1982, Cumming:1999} and implemented the algorithm as described in \citet{Zechmeister:2009b}. We chose a normalization such that 1 indicates a perfect fit of the data by a sine wave at a given period whereas 0 indicates no improvement compared to a fit of the data by a constant. To evaluate the false-alarm probability of any peak, we generated faked data set made of noise only. To make these virtual time series we used bootstrap randomization, i.e. we shuffled the original RVs and retained the date. Shuffling the RVs insures that no coherent signal is present in the virtual time series and keeping the dates conserve the sampling. For each trial we computed a periodogram and measured the power of the highest peak. With 10'000 trials we obtained a distribution of power maxima, which we used as a statistical description for the highest power one can expect if the periodogram was computed on data made of noise only. We searched for the power values that encompassed 68.3\%, 95.4\% and, 99.7\% of the distribution of power maxima (equivalent to 1-, 2-, and, 3-$\sigma$). A peak found with a power higher than those values (in a periodogram of the original time series) was hence attributed a FAP lower than 31.7, 4.6 or 0.3\%.

We started with a periodogram of the raw RVs. It shown a sharp peaks around periods $P=8.6$ and $1.13$ days (Fig.~\ref{fig1}, top panel). They have powers p=0.50 and 0.41, respectively, much above the power p=0.21 of a 0.3\% FAP. We noted they are both aliases of each other with our typical 1-day sampling and thus tried both periods as starting values  for a Keplerian fit. To perform the fit, we used a non-linear minimization with the Levenberg-Marquardt algorithm \citep{Press:1992}. We converged on local solutions with reduced $\chi^2$ (resp. rms) of 2.52$\pm$0.06 (resp. 4.53 m/s) and $3.02\pm0.06$ (resp. 5.02 m/s), respectively. We thus adopted $P_b=8.6$ day for the orbital period of the first planet.

We continued by subtracting the Keplerian orbit of planet 'b' to the raw RVs and by doing a periodogram of the residuals (Fig.~\ref{fig1}, second panel). We computed a power p=0.21 for the 0.3\% FAP threshold and located 8 peaks with more power. They had periods 0.996, 0.999, 1.002, 1.007, 1.038, 25.6, 227 and 625 day, and powers 0.48, 0.30, 0.30, 0.24, 0.30, 0.28, 0.25 and, 0.41, respectively. We identified that several candidates periods are aliases of each other and tried each as a starting value for a Keplerian fit, to a model now composed of two planets. We converged on local solutions with reduced $\chi^2$ (resp. rms) of 2.01 (resp. 3.55 m/s), $2.10$ (resp. 3.71 m/s), $1.98$ (resp. 3.50 m/s), $2.21$ (resp. 3.91 m/s), $2.13$ (resp. 3.76 m/s), $2.14$ (resp. 3.77 m/s), $2.19$ (resp. 3.87 m/s), $1.84$ (resp. 3.24 m/s), respectively. Among the peaks with highest significance, the one at P$\sim$600 day provided the best fit and we thus adopted that solution.

Next, we pursued the procedure and looked at the residuals around the 2-planet solution (Fig.~\ref{fig1}, third panel). We recovered some of the previous peaks, with even slightly more power excesses (p=0.30 and 0.28), at periods 25.6 and 1.038 day. We noted again that both periods are probably aliased of each other with the typical 1-day sampling. We performed 3-planet fit  trying both periods as initial guess for the third planet. We converged on $\chi^2=1.50$ (rms=2.59 m/s) and $\chi^2=1.53$ (rms=2.66 m/s) for guessed periods of 25.6 and 1.038 day, respectively. With the periodogram analysis, the solution with $P_b=25.6$ day is only marginally favored over the solution with $P_b=1.038$ day.

The fourth iteration unveiled one significant power excess around the period 1.006 day (p=0.22), as well as 2 other peaks above the 2-$\sigma$ confidence threshold, with periods 19.4 and 108 day (p=0.16 and 0.14 -- Fig.~\ref{fig1}, fourth panel). We note that the periods 1.006 and 108 day are each other aliases under our typical 1-day sampling. We tried all three periods (1.006, 19.4 and 108 days) as starting values and converged on $\chi^2=1.26$ (rms=2.15 m/s), $\chi^2=1.37$ (rms=2.32 m/s) and, $\chi^2=1.32$ (rms=2.26 m/s), respectively. Again no period is significantly favored. 

We adopted the solution with $P_d=108$ day and computed the periodogram of the residuals. The maximum power is seen again around 19.4 day, now above the 3-$\sigma$ confidence level. We checked that conversely, if we had adopted the solution with $P_d=19.4$ day, the period around 108 day (and 1.006 day) would now be the most significant, and above the $3-\sigma$ threshold too.

Eventually, the sixth iteration unveiled no additional significant power excess. The final 5-keplerian fit has a reduced $\chi^2=1.21$, for a rms=2.02 m/s. For reference, we give the orbital elements we derived in this section in Table~\ref{TabOrb2} (available in electronic form only).

\begin{figure}
\includegraphics[width=\linewidth]{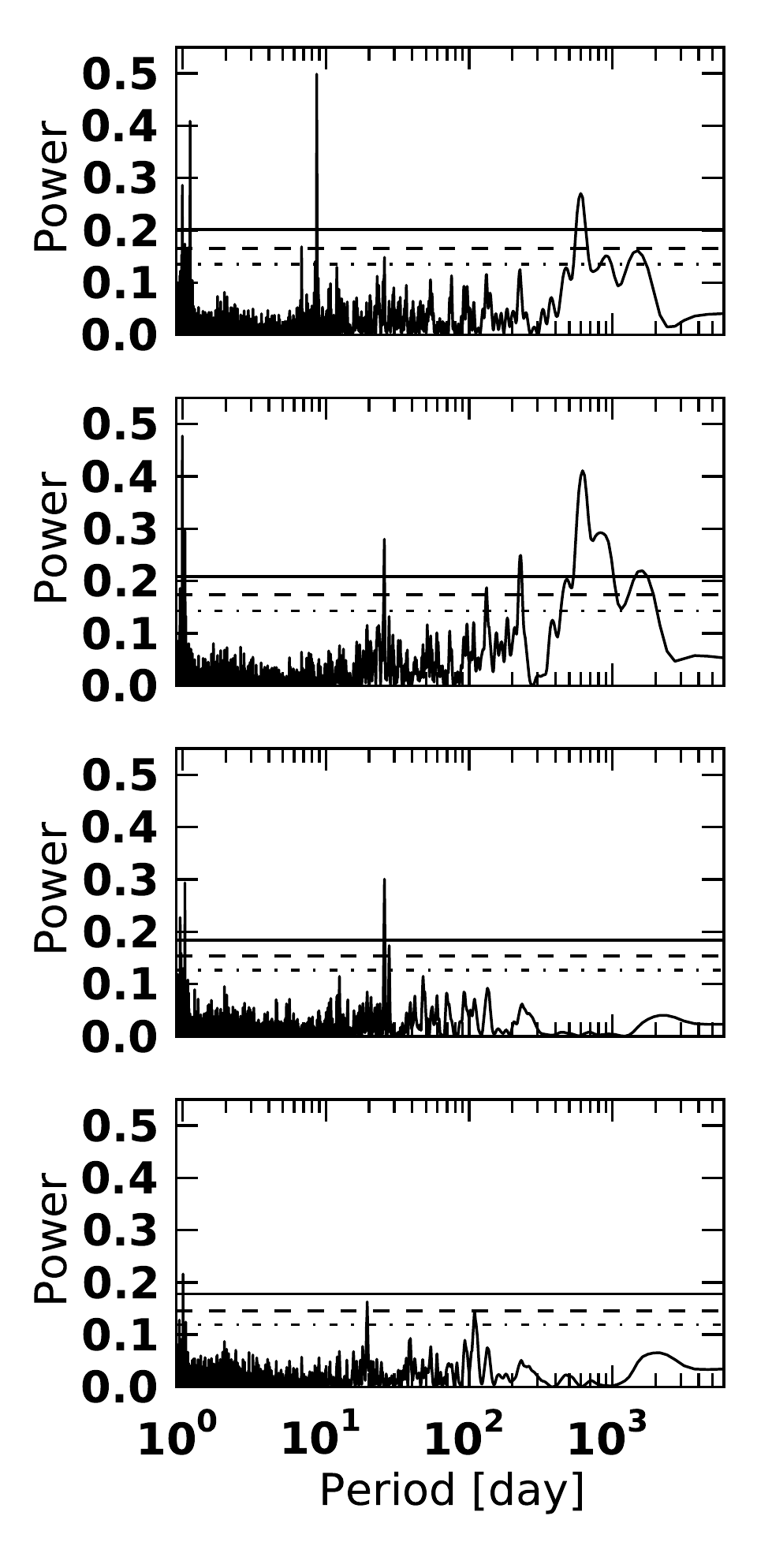}
\caption{Periodogram analysis of GJ\,163 RV time series, first 4 iterations. Horizontal lines mark powers corresponding to 31.7, 4.6 and 0.3\% false-alarm probability, respectively (i.e. equivalent to 1-, 2-, and 3-$\sigma$ detections).}
\label{fig1}
\end{figure}

\begin{figure}
\includegraphics[width=\linewidth]{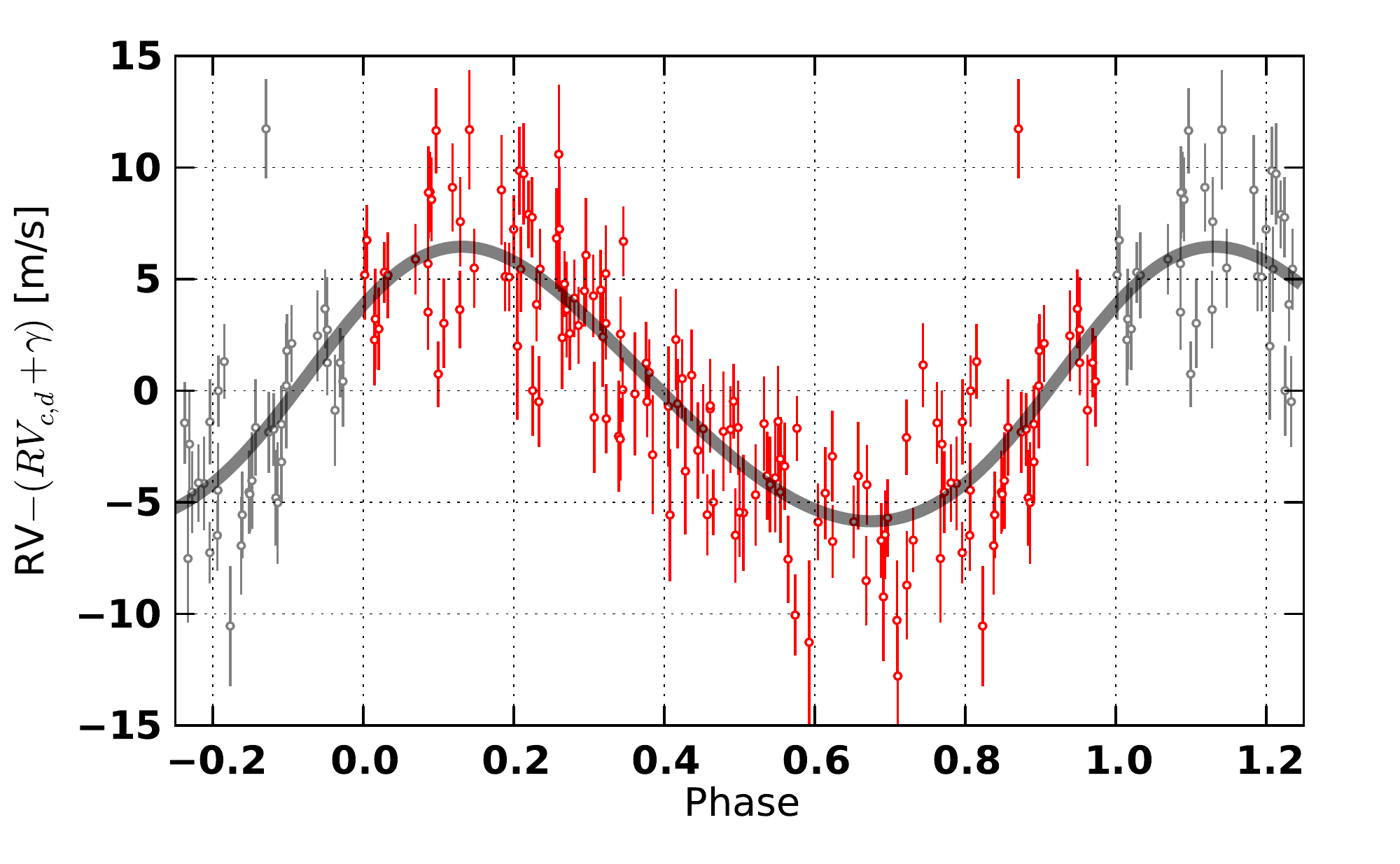}
\includegraphics[width=\linewidth]{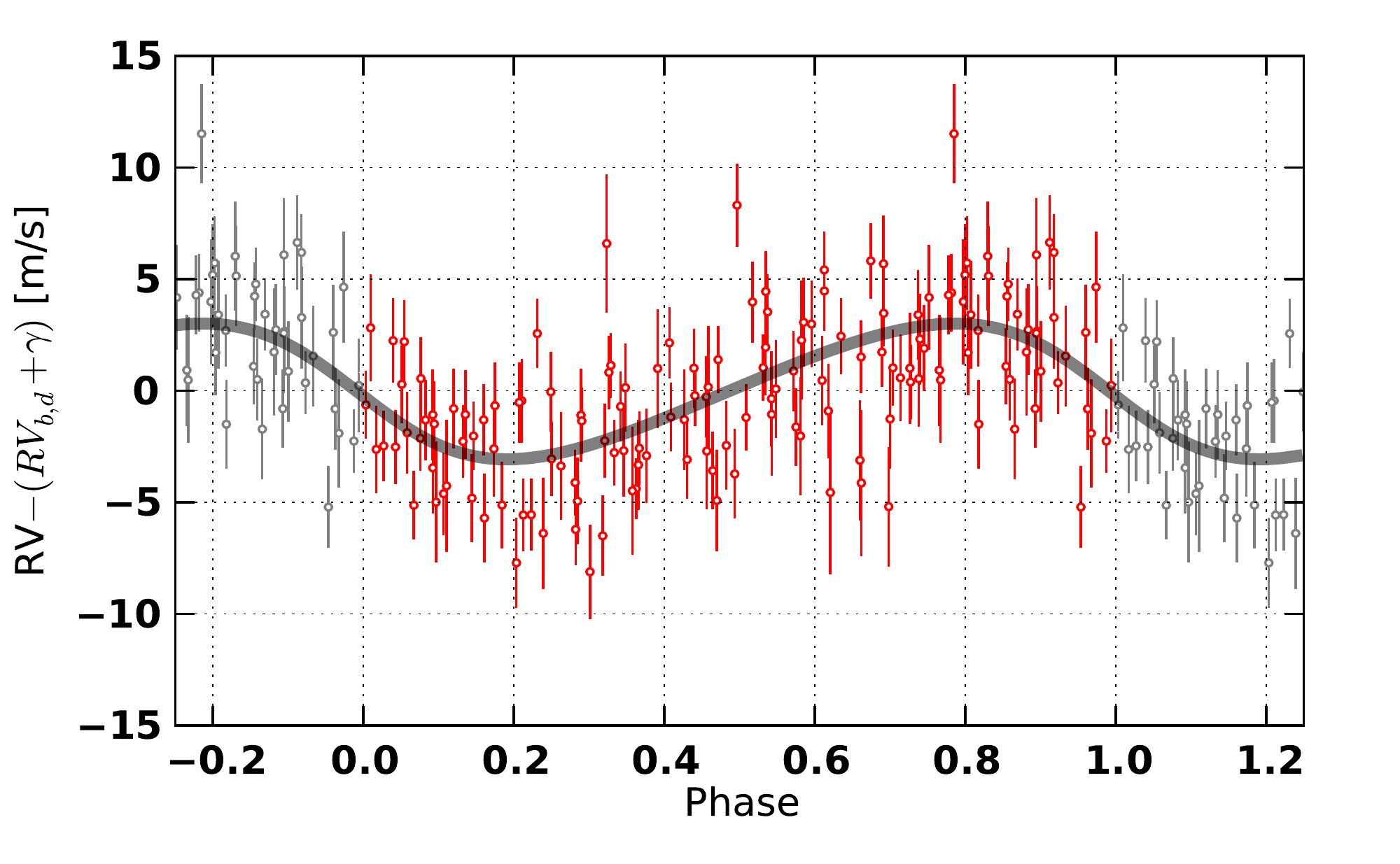}
\includegraphics[width=\linewidth]{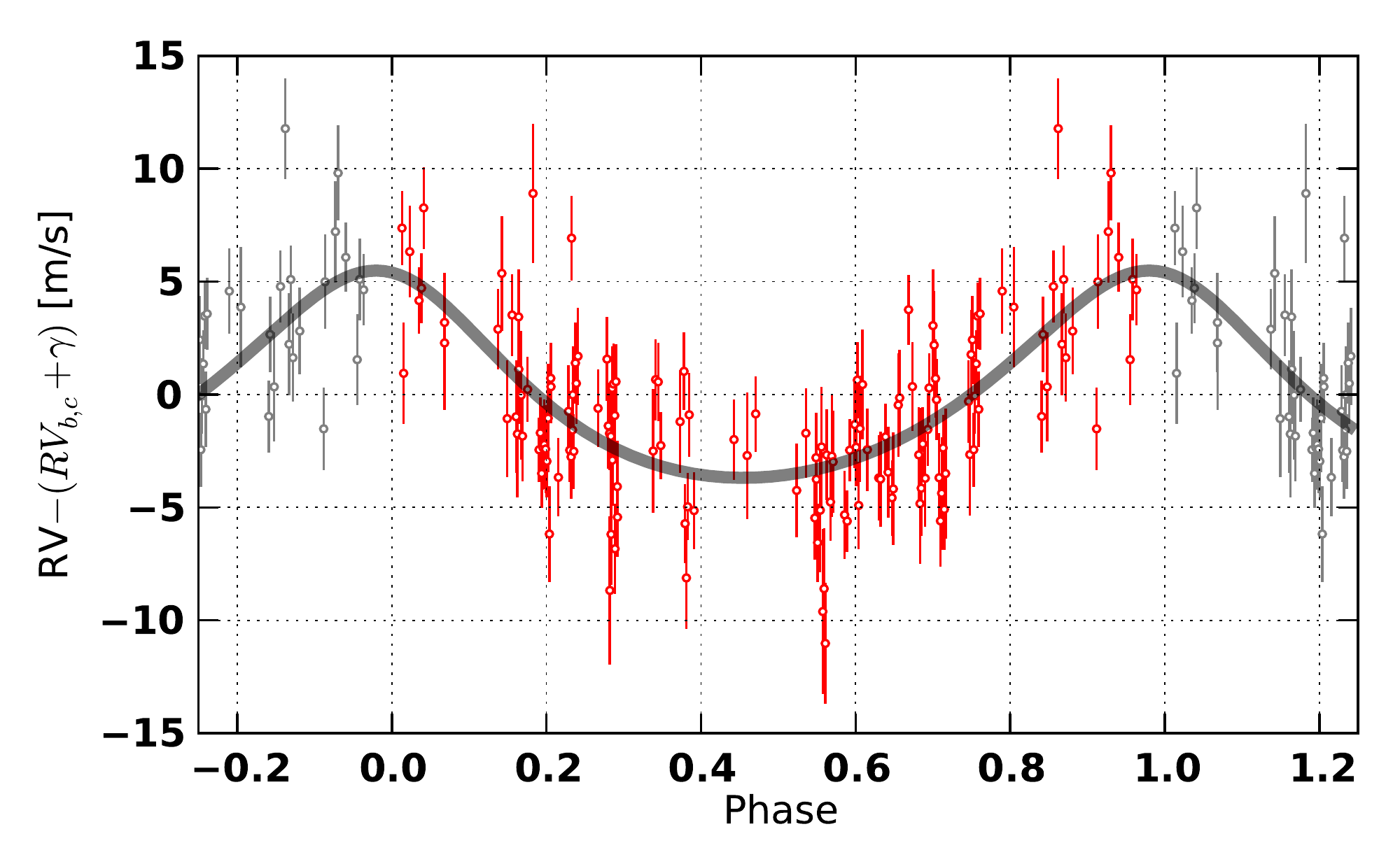}
\caption{Radial velocity curves for planets b, c and d, from top to bottom. }
\label{fig3}
\end{figure}

\section{\label{sect:activity}Challenging the planetary interpretation}

At this point, we identified up to 5 significant signals entangled in the RV data. If not caused by planets orbiting GJ\,163, some radial velocity periodic variations could be caused by stellar surface inhomogeneities such as plages or spots. The periodicity is then similar than the orbital period $P_{\rm rot}$, or might be one of its harmonic $P_{\rm rot}/2$, $P_{\rm rot}/3$, etc \citep{Boisse:2011}. Considering the activity of GJ\,163  (Sect.~\ref{sect:prop}), we found the rotation is moderate to long, likely greater than two more active stars of our sample, GJ\,176 and GJ\,674 (i.e. $P_{\rm rot}>35$ day), and possibly as long as the rotation period of GJ\,581 ($\sim$94 d). And therefore, up to three out of the five periodicities identified above might be confused with an activity-induced modulation: the 19.4, 25.6 and 108 day periodicities. In this section, we investigated time-variability of these signals (\S~\ref{subsec:fractioned}) and searched for their possible counterpart in various activity indicators (\S~\ref{subsec:indicators}).

\subsection{\label{subsec:fractioned}Search for changes in RV periodic signals}
To explore the possible non-stationarity of one signal, we fitted the data with a model composed of the 4 other signals and looked at the residuals. In practice, we chose to start the minimization close from the 5-planet solution. We used the solution with 5 planets (\S~\ref{subsect:periodo}) and removed from the solution the planet corresponding to the signal we want to study. We then performed a local minimization and computed the residuals, which thus include the signal of interest. Next, we divided the residual time-series in 3 observational seasons (2008, 2009, and 2010$+$2011). We did not included the observations before 2008 because there are too few and we grouped together 2010 and 2011 data. 

\begin{figure*}
\includegraphics[width=\linewidth]{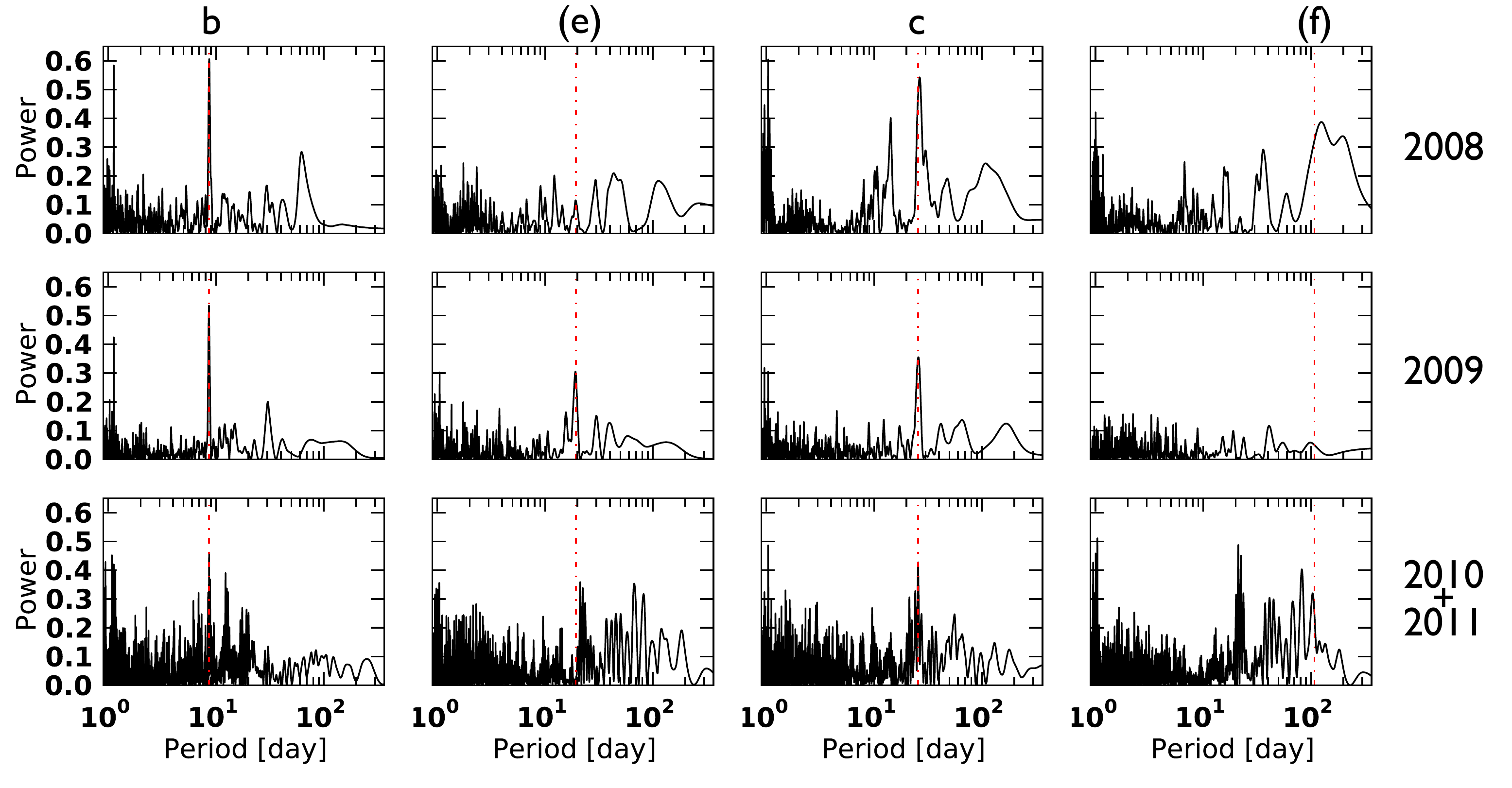}
\caption{Seasonal periodograms of residual time-series obtained after fitting the RV time-series with 4-planet models. From top to bottom, the rows are for seasons 2008, 2009, and 2010+2011, respectively. From left to right, the columns are periodograms to investigate signals $b$, $(e)$, $c$, and $(f)$, respectively. The periodicity of each signal is located with a vertical dashed red line. Power excesses are seen at all seasons for signals $b$ and $c$, but not for signals $(e)$ and $(f)$.}
\label{fig-splits}
\end{figure*}

We repeated the procedure for all signals but for the longest period (because the $\sim604$-day signal can not be recovered on the time-scale of one season). This produced 4$\times$3=12 peridograms, shown in Fig.~\ref{fig-splits}. To help locate where the unfitted signal should appear we located its period with a vertical red dashed line. 

For both signals $b$ and $c$, we see clear power excesses at the rights periods and for all seasons. This gives further credit that they are the result of orbiting planets. Conversely, the power excess expected for signal $(e)$ is seen in season 2009 only and no power excess is seen for signal $(f)$ in season 2009. This cast doubts on the nature of both signals $(e)$ and $(f)$ and call for more data before drawing further conclusions.

\subsection{\label{subsec:indicators}Periodicities in activity indicators}
Stellar activity can be diagnosed with spectral indices or by monitoring the shape of the spectral lines, both conveniently measured on the same spectra as those used to measure the radial velocities. We measured 2 spectral indices based on \ion{Ca}{ii} H \& K lines and on the H$\alpha$ line, as well as the full-width half maximum (FWHM) and the bisector span (BIS) of the cross-correlation function (CCF). Their values are given in Table~\ref{tab:rv} along with the radial-velocity measurements.

Among these indicators, we identified a significant periodicity for the FWHM only. Its periodogram indicates some power excess around a period of 30 days, with a false-alarm probability $<$0.3\% (i.e. a confidence level $>3\sigma$). We also looked for non stationarity in FWHM and found it is only pseudo-periodic. For instance, in 2008, the maximum power is seen at 30 days, with significant power around 19 day, compatible with the period $P_{(e)}$ identified in RV data. The possible link between this signal with the RV 19.4-day periodicity is however unclear since their strongest power is identified in periodograms of different seasons. We also show the periodogram of FWHM for the 2009 season, where the strongest pic is seen around the period 38 day (i.e. twice 19 day), albeit with a modest significance.

It is also unclear whether this stellar activity can be linked to the stellar rotation, as a $\sim$19-38 day rotational period would be short compared to our estimate in Sect.~\ref{sect:prop}.

\begin{figure}
\includegraphics[width=\linewidth]{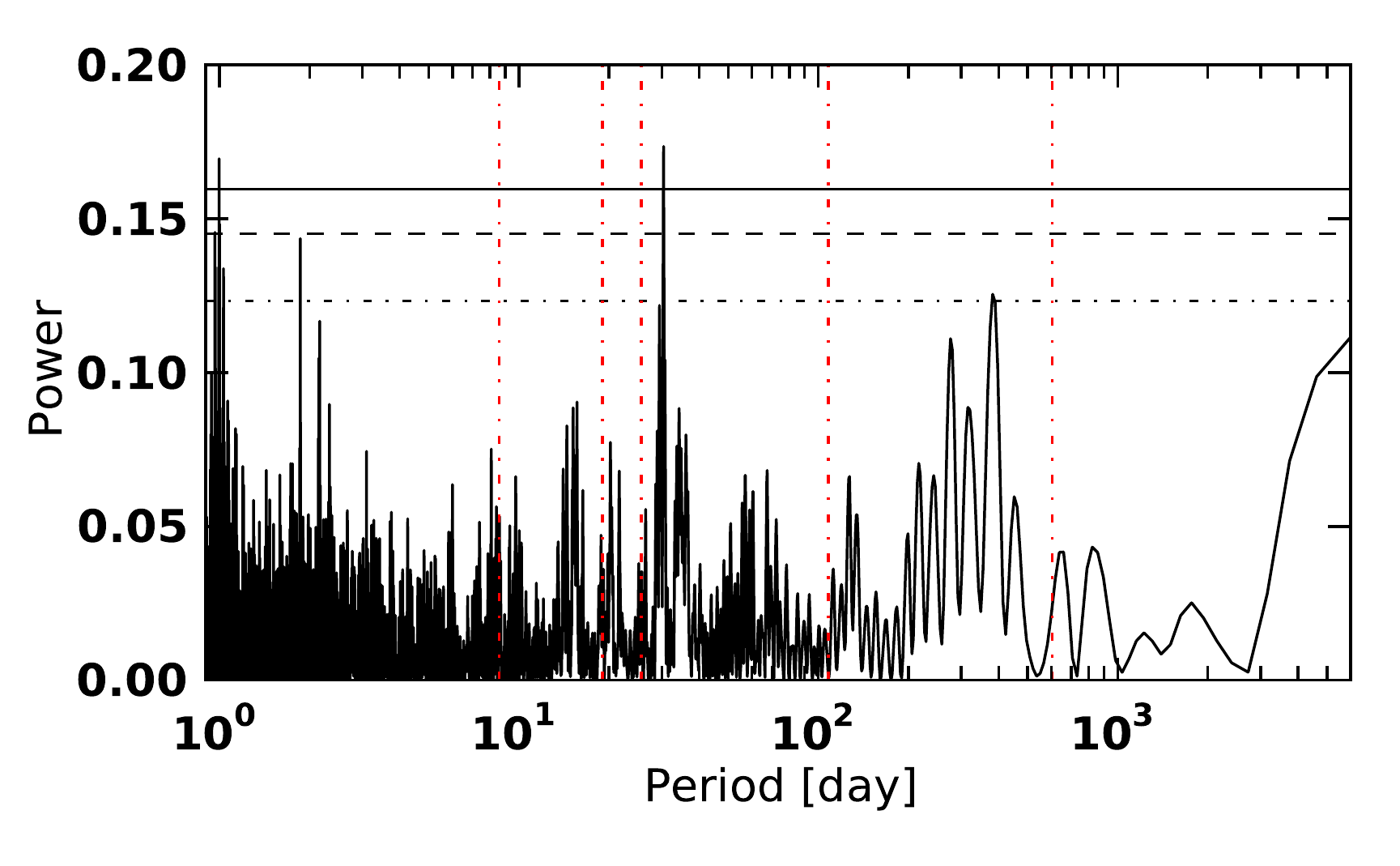}
\includegraphics[width=\linewidth]{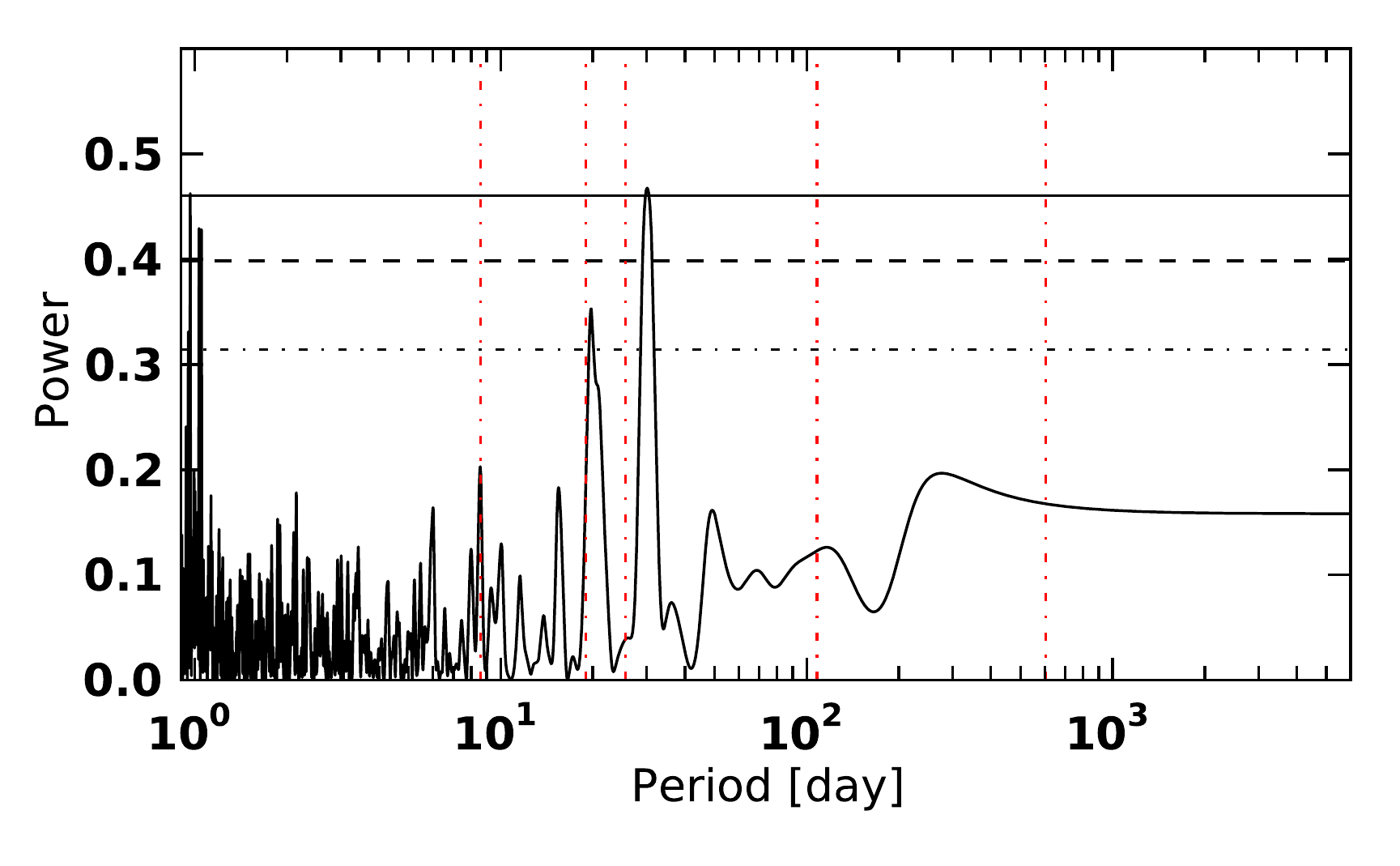}
\includegraphics[width=\linewidth]{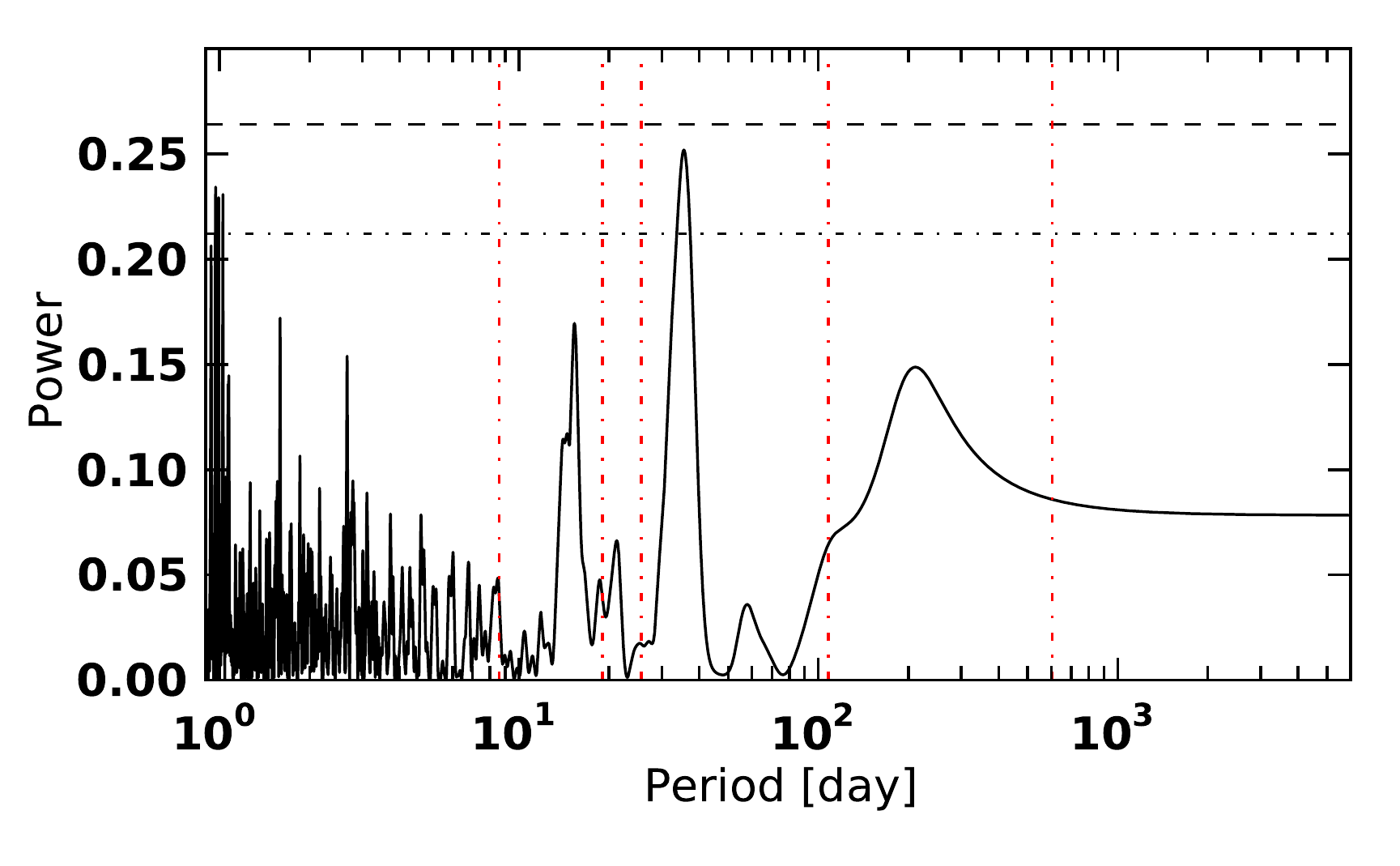}
\caption{Periodogram of the full-width half maximum of the cross-correlation function for both the whole data set (top panel), season 2008 only (middle panel) and season 2009 (bottom panel). For reference, the period of RV signals are shown with vertical red dashed lines.}
\label{fig-fwhm}
\end{figure}

\begin{table*}
\caption{
\label{table:fit}
Modeled and inferred parameters for GJ~163 system. 
}
\begin{tabular}{l@{}lcc}
\hline
                              & {\bf Unit}  & {\bf Prior} & {\bf Posterior}  \\
\hline
Systemic velocity, $\gamma$       & [km/s]                 & Uniform ($\gamma_{\rm min}=-100$, $\gamma_{\rm max}=+100$) & 58.59728 $\pm 0.00026$\\\\
Orbital period, $P_b$                 & [day]                    & Jeffreys ($P_{\rm min}=1$, $P_{\rm max}=10^4$) & 8.63182 $\pm 0.00155$ \\
Radial-velocity semi-amplitude, $K_b$ ~~~~           & [m/s]              & Jeffreys ($K_{\rm min}=0.1$,$K_{\rm max}=100$)     &6.13 $\pm 0.33$ \\
Orbital eccentricity, $e_b$                 &  [~]                          & Uniform ($e_{\rm min}=0$, $e_{\rm max}=0.91$) & 0.073 $\pm$ 0.050; $<0.101$  (1-$\sigma$ upper limit)\\
Argument of periastron, $\omega_b$         &       [rad]                 & Uniform ($\omega_{\rm min}=0$, $\omega_{\rm max}=2\pi$)   & 1.234 $\pm$ 0.953 \\
Mean longitude, $\lambda_b$         &       [rad]                 &   & 0.111 $\pm$ 0.274 \\
Semi-major axis, a$_b$ &  [AU]                    && 0.06070 $\pm 0.00001$\\
Time of inferior conjunction, $T_{tr,b}$ &  [day]                    && 52936.1209 $\pm 0.4038$\\
Planetary minimum mass, m$_b \sin i$                 & [M$_\oplus$]    && 10.6 $\pm 0.6$ \\\\
Orbital period, $P_c$                 & [day]                    & Jeffreys ($P_{\rm min}=1$, $P_{\rm max}=10^4$) & 25.63058 $\pm 0.02550$ \\
Radial-velocity semi-amplitude, $K_c$ ~~~~           & [m/s]              & Jeffreys ($K_{\rm min}=0.1$,$K_{\rm max}=100$)     &2.75 $\pm 0.35$ \\
Orbital eccentricity, $e_c$                 &  [~]                          & Uniform ($e_{\rm min}=0$, $e_{\rm max}=0.91$) & 0.099 $\pm$ 0.086; $<0.144$  (1-$\sigma$ upper limit)\\
Argument of periastron, $\omega_c$         &       [rad]                 & Uniform ($\omega_{\rm min}=0$, $\omega_{\rm max}=2\pi$)   & 3.962 $\pm$ 1.394 \\
Mean longitude, $\lambda_c$         &       [rad]                 &   & 0.428 $\pm$ 0.524 \\
Semi-major axis, $a_c$ &  [AU]                    && 0.1254 $\pm 0.0001$\\
Time of inferior conjunction, $T_{tr,c}$ &  [day]                    && 52922.2303 $\pm 2.2951$\\
Planetary minimum mass, m$_c \sin i$                 & [M$_\oplus$]    && 6.8 $\pm 0.9$ \\\\
Orbital period, $P_d$                 & [day]                    & Jeffreys ($P_{\rm min}=1$, $P_{\rm max}=10^4$) & 603.95116 $\pm 7.55862$ \\
Radial-velocity semi-amplitude, $K_d$ ~~~~           & [m/s]              & Jeffreys ($K_{\rm min}=0.1$,$K_{\rm max}=100$)     &4.42 $\pm 0.51$ \\
Orbital eccentricity, $e_d$                 &  [~]                          & Uniform ($e_{\rm min}=0$, $e_{\rm max}=0.91$) & 0.373 $\pm$ 0.077 \\
Argument of periastron, $\omega_d$         &       [rad]                 & Uniform ($\omega_{\rm min}=0$, $\omega_{\rm max}=2\pi$)   & 2.064 $\pm$ 0.357 \\
Mean longitude, $\lambda_d$         &       [rad]                 &   & 2.530 $\pm$ 0.301 \\
Semi-major axis, $a_d$ &  [AU]                    && 1.0304 $\pm 0.0086$\\
Time of inferior conjunction, $T_{tr,d}$ &  [day]                    && 52876.6622 $\pm 35.8448$\\
Planetary minimum mass, m$_d \sin i$                 & [M$_\oplus$]    && 29.4 $\pm 2.9$ \\
\hline
\end{tabular}\\
$T_{\rm epoch}$ = 52942.80392 day\\
An additional 10\% uncertainty should be added quadratically to the planetary mass uncertainties when accounting for the 10\% uncertainty on the stellar mass.
\end{table*}

%______________________________________________________________
%
\section{\label{sect:dyn}Dynamical analysis}
%______________________________________________________________

After analyzing the RV data with both a MCMC algorithm and iterative periodograms, we identified up to 5 superimposed coherent signals. In Sect.~\ref{sect:activity} we scrutinized several activity indicators and looked for non-stationarity of these signals to finally cast doubts on the planetary nature for two of them. We retained a nominal solution with 3 planets (Table\,\ref{table:fit}) and now perform a dynamical analysis.

The orbital solution given in Table\,\ref{table:fit}, shows a planetary
system composed of three planets, two of them in very tight orbits ($a_b = 0.06 $ and $a_c = 0.13$~AU), and another further away, but in an eccentric orbit, such that the minimum distance at pericentre is only 0.65~AU.
The stability of this system is not straightforward, in particular taking into
account that the minimum masses of the planets are of the same order as
Neptune's mass.
As a consequence, mutual gravitational interactions between planets in the
GJ\,163 system cannot be neglected and may give rise to some instability.

\subsection{Secular coupling}

\begin{table}
 \caption{Fundamental frequencies for the nominal orbital solution in
 Table\,\ref{table:fit}. 
 \label{Tdyn1}} 
 \begin{center}
 \begin{tabular}{crrr}
 \hline\hline
      & Frequency   & Period & Angle \\
      & ($^\circ/yr$) & (yr) & (deg) \\
 \hline
 $n_b$ & 15231.673258 &             0.023635 &       6.1715 \\
 $n_c$ &   5134.634175 &              0.070112 &     24.7197 \\
 $n_d$ &     217.437041 &              1.655652 &   144.9791 \\
 $g_1$ &          0.054525 &       6602.468907 & -178.2139 \\
 $g_2$ &          0.243159 &       1480.513345 & -122.6748 \\
 $g_3$ &          0.000241 & 1490684.180450 & 118.3562 \\ \hline
\end{tabular}
\end{center}
$n_b$, $n_c$ and $n_d$ are the mean motions, and $g_1$, $g_2$ and $g_3$ are the
 secular frequencies of the pericentres.
\end{table}

The ratio between the orbital periods of the two innermost planets determined 
by the fitting process (Table\,\ref{table:fit}) is $ P_c
/ P_b = 2.97 $, suggesting that the system may be trapped in a 3:1
mean motion resonance.
To test the accuracy of this scenario, we performed a frequency analysis of the 
nominal orbital solution listed in Table\,\ref{table:fit} computed over 1~Myr.
The orbits of the planets are integrated with the 
symplectic integrator SABA4 of \citet{Laskar_Robutel_2001}, using a step size
of 0.01~yr, including general relativity corrections.
We conclude that, in spite of the proximity of the 3:1 mean motion resonance, when we adopt the minimum values for the masses, the two planets in the GJ\,163 system are not trapped in this resonance.

\begin{figure}
    \includegraphics*[width=\linewidth,angle=0]{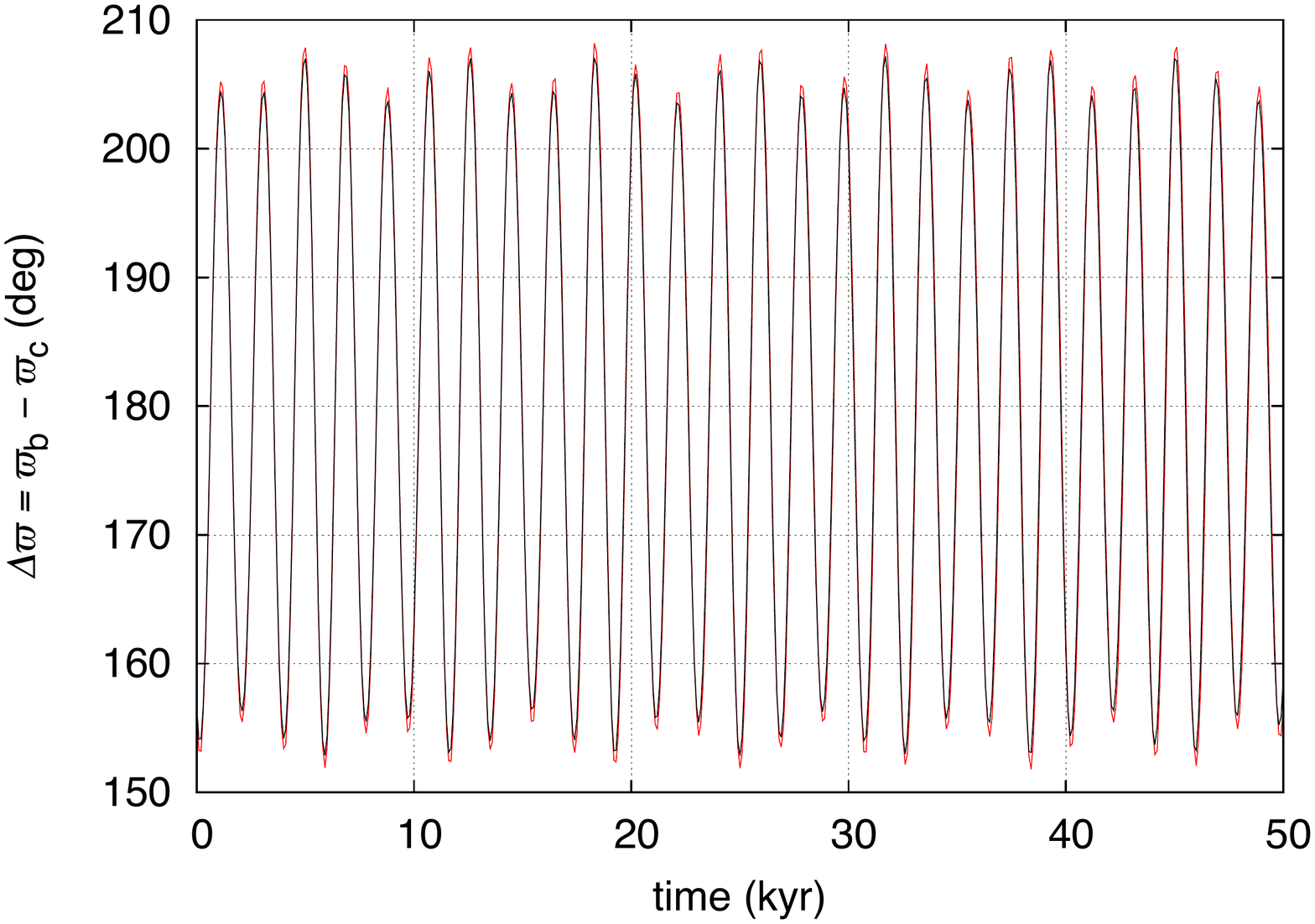} 
  \caption{Evolution of the angle  $\Delta \varpi = \varpi_b-\varpi_c$ (red line) that oscillates around $180^\circ$ with a maximal amplitude of $28^\circ$.  The black line also gives the $\Delta \varpi $ evolution, but obtained with the linear secular model (Eqs.\,\ref{eq.lape}).
   \label{Fdyn1}   }
\end{figure}

The fundamental frequencies of the systems are then
the mean motions $n_b$, $n_c$ and $n_d$, and the three secular frequencies of the 
pericentres $g_1$, $g_2$ and $g_3$ (Table\,\ref{Tdyn1}).
Because of the proximity of the two innermost orbits, there is  a strong  coupling 
within the secular system \citep[see][]{Laskar_1990}. Both planets $b$ and $c$ precess with  the same precession frequency $g_2$, which has a period of 1480~yr. 
The two pericentre are thus locked and $\Delta \varpi = \varpi_c-\varpi_b$
oscillates around $180^\circ$, with a maximal amplitude of about  $28^\circ$ (Fig.\,\ref{Fdyn1}).
This behavior is not  a dynamical resonance, but merely the result of the linear secular coupling. 

To present the solution in a clearer way, it is useful to make a linear change of variables into eccentricity proper modes \citep[see][]{Laskar_1990}. 
In the present case, due to the proximity of the 3:1 mean motion resonance and due to the high value of the outer planet eccentricity, the linear transformation is numerically 
obtained by a frequency analysis of the solutions.
Using the classical complex notation,   
\begin{equation}
z_\k = e_\k \mathrm{e}^{i \varpi_\k} \ ,
\end{equation}
for $\k = b, c, d$, we have for the linear Laplace-Lagrange solution
\begin{equation}
\left(\begin{array}{c} z_b\\ z_c \\z_d \end{array}\right)= 
\SS
\left(\begin{array}{c} u_1\\ u_2 \\u_3 \end{array}\right) \ ,
\label{eq.lape}
\end{equation}
where $\SS$ is given by
\begin{equation}
\SS = 
\left(\begin{array}{rrr} 
 0.019139, &  -0.080497, & \phantom{-}0.001192 \\
 0.018850, &   0.087791, & 0.001538 \\
-0.000015, & -0.000011, & 0.373356 \\
\end{array}\right)  \ .
\label{eq.lape2}
\end{equation}
The proper modes $u_\l$ (with $\l = 1, 2, 3$) are obtained from the $z_\k$ by inverting the above linear relation. 
To good approximation, we have $u_\l \approx  \mathrm{e}^{i (g_\l t +\phi_\l)}$, where $g_\l$ and $\phi_\l$ are given in Table\,\ref{Tdyn1}.

From Eqs.\,\ref{eq.lape}, it is then easy to understand the meaning of the observed 
libration between the pericentres $\varpi_b$ and $\varpi_c$. 
Indeed, for both planets  $b$ and $c$, the dominant term is $u_2$  with frequency $g_2$, and they thus both precess with an average value of $g_2$ (black line, Fig.\,\ref{Fdyn1}). 

\begin{figure}
    \includegraphics*[width=\linewidth]{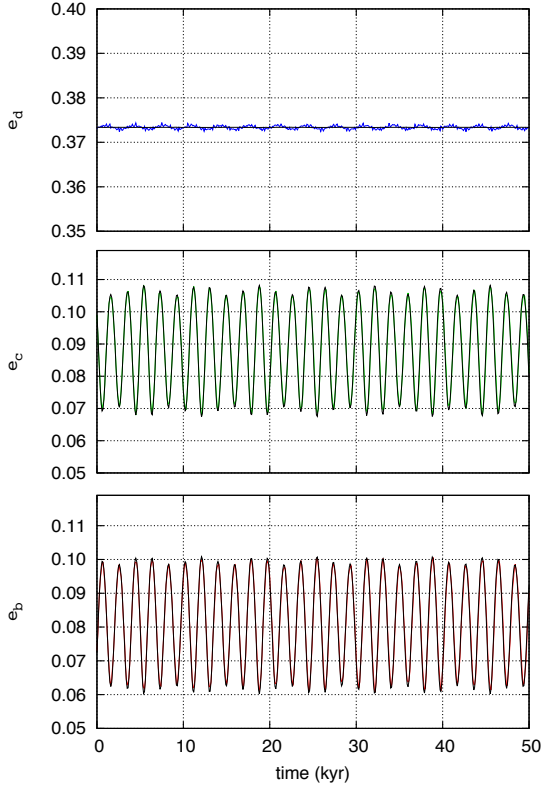} \\    
  \caption{Evolution of the GJ\,163 eccentricities with time, starting with the orbital solution from Table\,\ref{table:fit}. The color lines are the complete solutions for the various planets (b: red, c: green,  d: blue), while the black curves are the associated values obtained with the linear secular model (Eqs.\,\ref{eq.lape}). 
   \label{Fdyn2}}   
\end{figure}

It should also be noted that Eqs.\,\ref{eq.lape} provide good approximations of the long-term evolution of the eccentricities. 
In Figure~\,\ref{Fdyn2} we plot the eccentricity evolution
with initial conditions from Table \ref{table:fit}. 
Simultaneously, we plot with black lines the evolution of the same elements given by the above secular, linear approximation.  
The eccentricity variations are very limited and described well by the secular approximation. 
The eccentricity of planets $b$ and $c$ are within the ranges $ 0.061 < e_b < 0.101 $ and $
0.067 < e_c < 0.109 $, respectively. These variations are  
driven mostly by  the  secular frequency $g_2$, of period approximately $ 1480 $~yr (Table\,\ref{Tdyn1}). 
The eccentricity of planet $d$ is nearly constant with $ 0.372 < e_d < 0.374 $ (Fig.\,\ref{Fdyn2}).

\subsection{Stability analysis}

\begin{figure}
  \centering
    \includegraphics*[width=8.8cm]{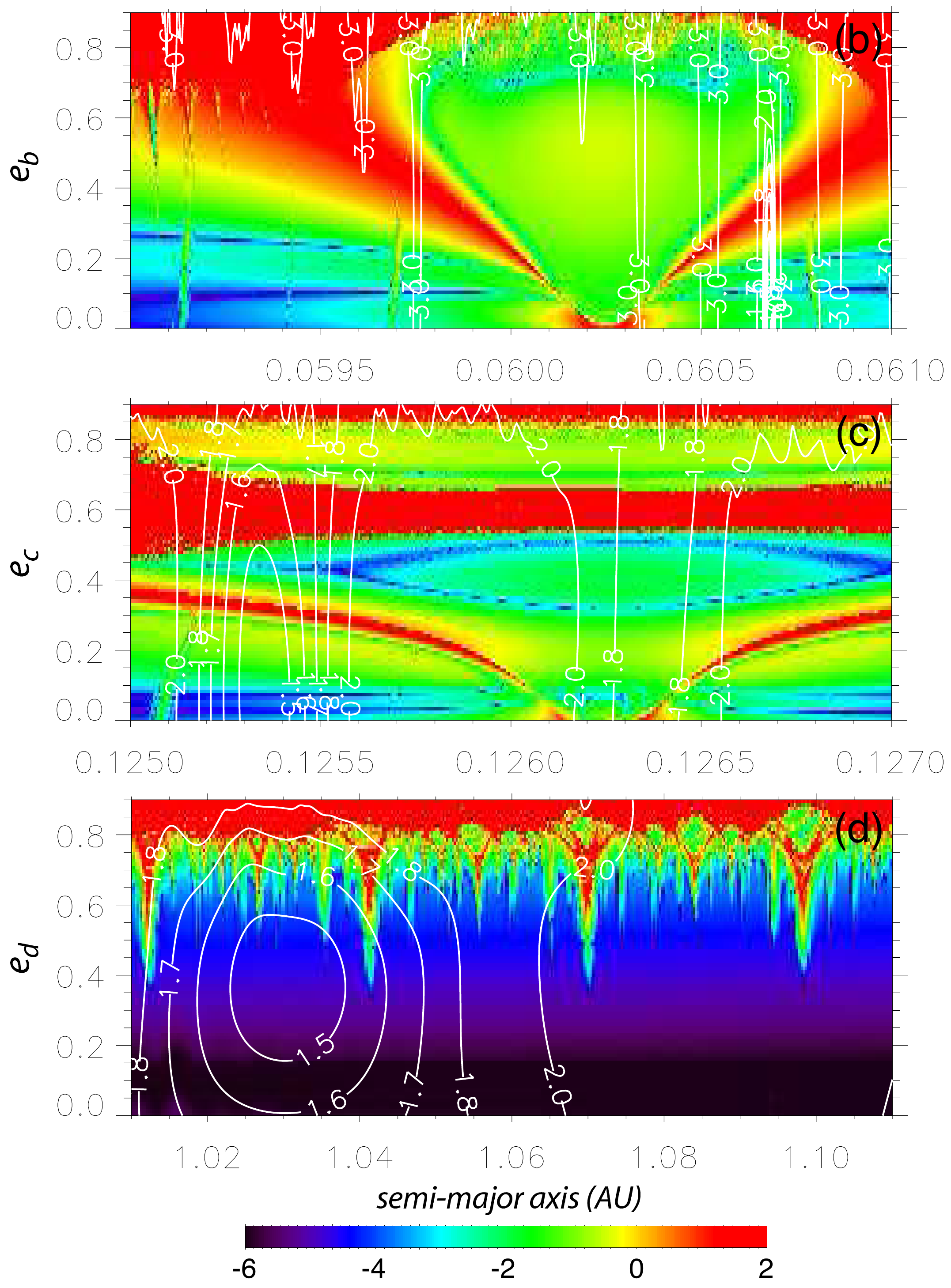} 
  \caption{
  Stability analysis of the nominal fit (Table\,\ref{table:fit}) 
  of the {GJ}\,163 planetary system. For fixed initial conditions, the phase space of the 
  system is explored by varying the semi-major axis $a_\k$ and eccentricity 
  $e_\k$ of each planet, respectively $b$, $c$, and $d$. The step size is $10^{-5}$~AU in semi-major axis and $10^{-2}$ in eccentricity. 
  For each initial condition, the system is integrated over 200~yr
  and a stability criterion is derived  with the frequency analysis of the mean longitude
  \citep{Laskar_1990,Laskar_1993PD}.
  As in \citet{Correia_etal_2005,Correia_etal_2009,Correia_etal_2010}, the chaotic diffusion 
  is measured by the variation in the frequencies. The ``red'' zone corresponds to highly unstable 
  orbits, while  the ``dark blue'' region can be assumed to be stable on a
  billion-years timescale.
  The contour curves indicate the value of $\chi^2$ obtained for each choice of parameters.
  \label{Fdyn3}}   
\end{figure}

In order to analyze the stability of the nominal solution (Table\,\ref{table:fit}) and
confirm that the inner subsystem is outside of the 3:1 mean motion resonance, 
we performed a global frequency analysis \citep{Laskar_1993PD} 
in the vicinity of this solution, in the same way as
achieved for other planetary systems \citep[e.g.][]{Correia_etal_2005,Correia_etal_2009,Correia_etal_2010}.

For each planet, the system is integrated on a regular 2D mesh of initial conditions, 
with varying semi-major axis and eccentricity, while the other parameters are 
retained at their nominal values (Table\,\ref{table:fit}). 
The solution is integrated over 200~yr for each initial condition and a stability indicator is computed to be the variation in the measured mean motion over the two consecutive 
100~yr intervals of time \citep[for more details see][]{Correia_etal_2005}.
For regular motion, there is no significant variation in the mean motion along the trajectory, while it can vary significantly for chaotic trajectories. 
The result is reported using a color index in Figure~\ref{Fdyn3}, where ``red'' represents the strongly chaotic trajectories, and ``dark blue''  the extremely stable ones. 

In Figure\,\ref{Fdyn3} we show the wide vicinity of the best fitted solution, the minima of the $\chi^2$ level curves corresponding to the nominal parameters (Table\,\ref{table:fit}).
For the inner system ({\it top} and {\it center} pannels) we observe the presence of the large 3:1 mean motion resonance. 
We confirm that the present system is outside the 3:1 resonance, in a more stable area at the bottom right side (Fig.\,\ref{Fdyn3}, {\it top}), or at the bottom left side (Fig.\,\ref{Fdyn3}, {\it center}).
These results are somehow surprising, because if the system had been previously captured inside the 3:1 mean motion resonance, we would expect that the subsequent evolution drive it to the opposite side of it, where the period ratio is above 3, instead of 2.97. 
Indeed, during the initial stages of planetary systems, capture in mean motion resonances can occur, as a result of orbital migration due to the interactions within a primordial disk of planetesimals,  \citep[e.g.][]{Papaloizou_2011}.
However, as the eccentricities of the planets are damped by tidal interactions with the star, this equilibrium becomes unstable.
For first order mean motion resonances it has been demonstrated that the system exits the resonance with a higher period ratio \citep[][]{Lithwick:2012, Delisle_etal_2012, Batygin:2013}, and this behavior should not differ much for higher order resonances.

For the outer planet (Fig.\,\ref{Fdyn3}, {\it bottom}), we observe that, although the planet lies in a very stable region. 
Nevertheless, since the contour curves of minimal $\chi^2$ vary smoothly is this zone (contrarily to those for the inner system), we conclude that this eccentricity may be overestimated.
Additional observational data will help to solve this issue, since longer orbital periods become better determined as we acquire data for extended time spans (because we cover more revolutions of the planet around the star).
Since the system is already stable with the nominal parameters from Table~\ref{table:fit}, we do not explore more deeply this possibility in the present paper, but more detailed dynamical studies on this system must take this possibility into account.

We also tested shortly the stability of the 5-planet solution (Table~\ref{TabOrb2}) and found it is not stable (even with eccentricities of planets $e$ and $f$ fixed to zero), in particular due to planet $e$.

\subsection{Long-term orbital evolution}

\begin{figure}
    \includegraphics*[height=8.8cm,angle=270]{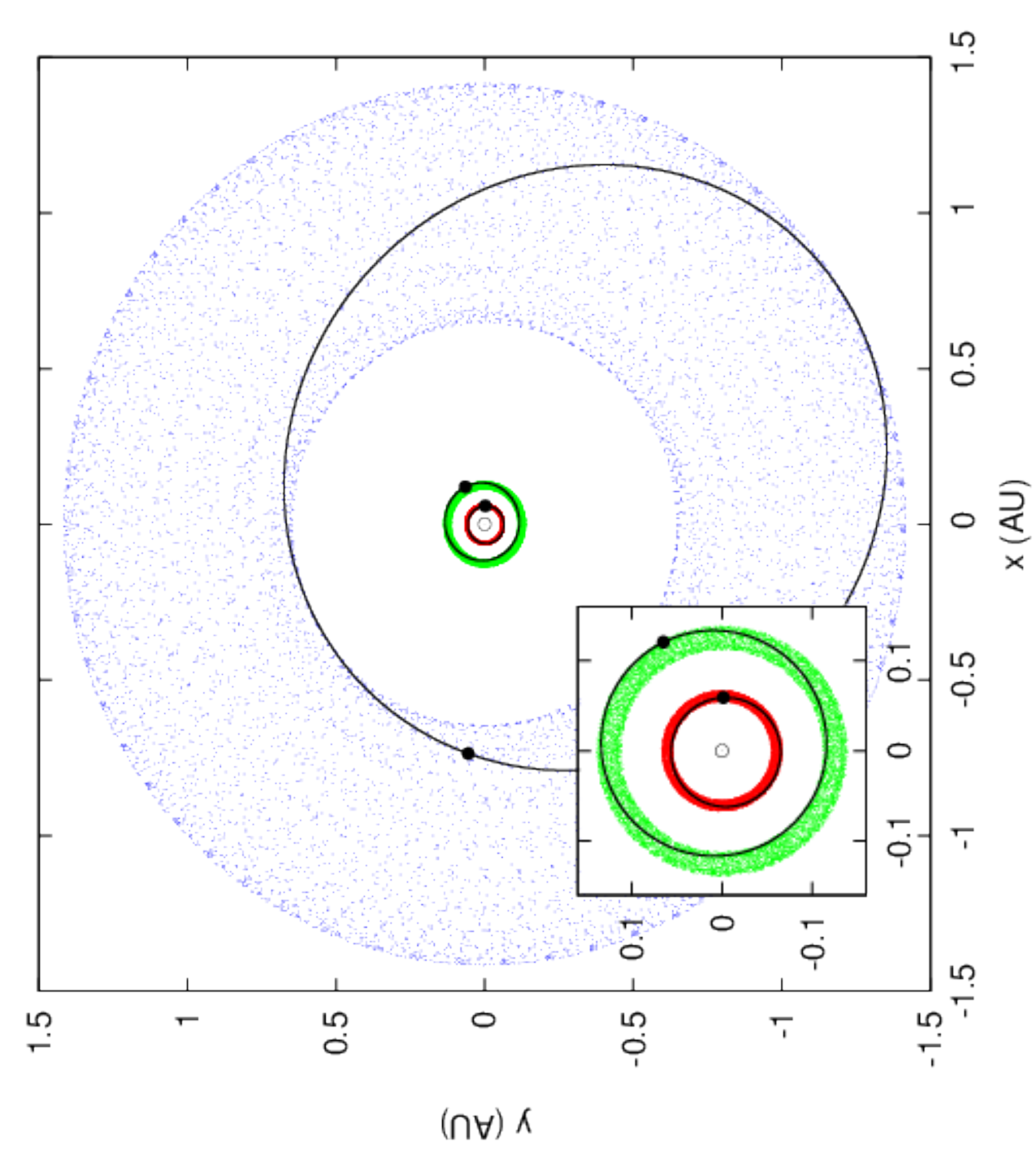} 
  \caption{Long-term evolution of the GJ\,163 planetary system over 1~Gyr
  starting with the orbital solution from Table\,\ref{table:fit}. We did not include
  tidal effects in this simulation. 
  The panel shows a face-on view of the system invariant plane. $x$ and $y$ are spatial
  coordinates in a frame centered on the star. Present orbital solutions are traced
  with solid lines and each dot corresponds to the position of the planet every
  0.1~Myr. The semi-major axes are almost constant, and the eccentricities
  present slight variations ( $ 0.061 < e_b < 0.101 $, $ 0.067 < e_c < 0.109 $, and $ 0.372 < e_d < 0.374 $).  \label{Fdyn6}}   
\end{figure}

\begin{figure*}
    \includegraphics*[width=18cm]{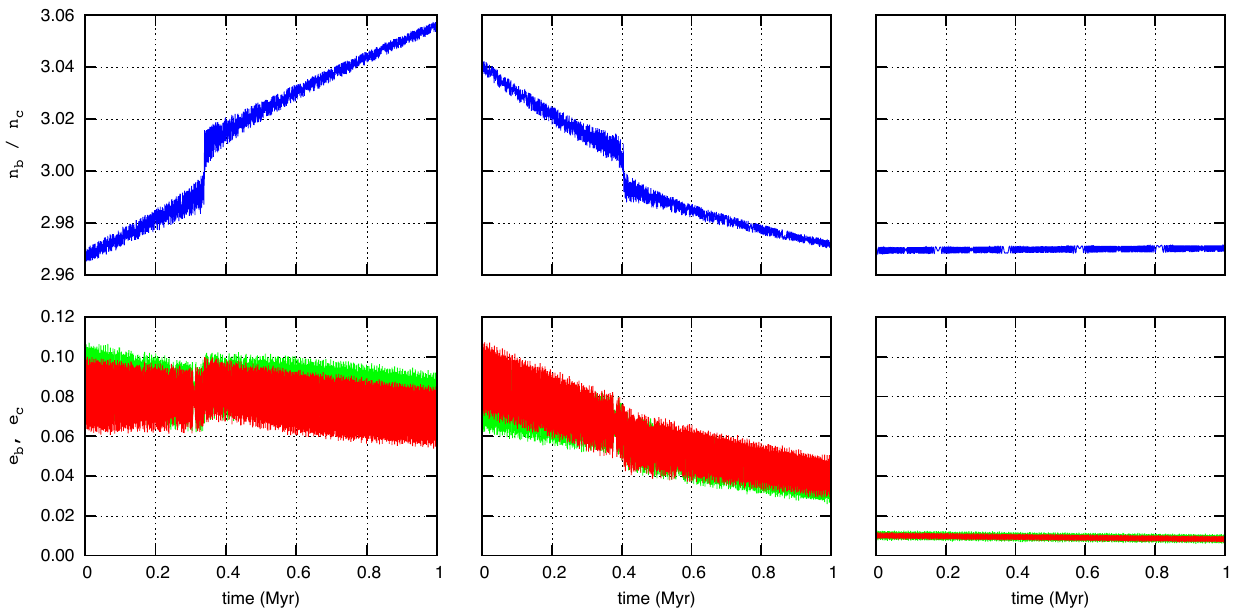} 
  \caption{Some possibilities for the long-term evolution of the GJ\,163 planetary system over 1~Myr, including tidal effects with $\Delta t_\k = 10^5$\,s.
  Time scales are inversely proportional to $\Delta t$ (Eq.\ref{eq.tides}), so 1~Myr of evolution roughly corresponds to 1~Gyr with $\Delta t_\k = 100$\,s ($Q_\k  \sim 10^3$) or 10~Gyr with $\Delta t_\k = 10$\,s ($Q_\k  \sim 10^4$). We show the ratio $P_c / P_b $ of the orbital periods of the two inner planets ({\it top}) and their eccentricities $e_b$ (red) and $e_c$ (green) ({\it bottom}). We use three different sets of initial conditions: Table\,\ref{table:fit} ({\it left}); 
Table\,\ref{table:fit} with $a_c =  0.060679$ and $\Delta t_c = 5 \times 10^7 $\,s ({\it middle});  
Table\,\ref{tidestab} ({\it right}).  \label{Fdyn7}}   
\end{figure*}

From the previous stability analysis, it is clear that the 
 GJ\,163 planetary system listed in Table\,\ref{table:fit} is stable over Gyr timescale.
Nevertheless, we also tested directly this by performing a numerical integration
of the orbits.

In a first experiment, we integrate the system over 1~Gyr using the symplectic integrator SABA4 of
\citet{Laskar_Robutel_2001} with a step size of 0.01~yr, including general relativity corrections, but without tidal effects.
The result displayed in Figure~\ref{Fdyn6} show that the orbits indeed 
evolve in a regular way, and remain stable throughout the simulation,
which is of the same order as the age of the star. 

Since the two inner planets are very close to the star, in a second experiment we run a numerical simulation that includes tidal effects.
Several tidal models have been developed so far, from the simplest ones to the more complex \citep[for a review see][]{Correia_etal_2003, Efroimsky_Williams_2009}.
The qualitative conclusions are more or less unaffected, so, for simplicity, we
adopt here a linear model with constant $\Delta t$ \citep{Singer_1968}, 
where $\Delta t$ is a time delay between the initial perturbation of the planet and the maximal tidal deformation.
The tidal force acting on each planet is then given by
\citep[][]{Mignard_1979}: 
\begin{equation}
\vec{F}_\k = - \Delta t_\k \frac{3 k_2 G M^2 R_\k^5}{r_\k^{10}} \big( 2 (\vec{r}_\k \cdot \dot{\vec{r}_\k} ) \vec{r}_\k + r_\k^2 (\vec{r}_\k \times \vec{\omega}_\k + \dot{\vec{r}_\k} ) \big) \ , \label{eq.tides}
\end{equation}
where $\vec{r}_\k$ is the position of each planet relative to the star, $k_2 $ is the potential Love number, $G$ is the gravitational constant, $M$ is the mass of the star, $R_\k$ is the planet radius, and $\vec{\omega}_\k $ is the spin vector of the planet.
Because the spin evolves in a much shorter timescale than the orbit \citep[e.g.][]{Correia_2009}, we consider that the spin axis is normal to the orbit, and its norm is given by the equilibrium rotation for a given eccentricity \citep[Eq.\,48,][]{Correia_etal_2011}: 
\begin{equation}
\frac{\vec{\omega}_\k}{n_\k} = \frac{(1 + \frac{15}{2}e_\k^2 + \frac{45}{8}e_\k^4 + \frac{5}{16}e_\k^6)}{(1 + 3e_\k^2 + \frac{3}{8}e_\k^4) (1-e_\k^2)^{3/2}} \hat{\vec{s}}_\k 
\ , \quad 
\hat{\vec{s}}_\k = \frac{\vec{r}_\k \times \dot{\vec{r}_\k}}{|| \vec{r}_\k \times \dot{\vec{r}_\k} ||} \ .
\end{equation}

In this experiment we use the ODEX integrator \citep[e.g.][]{Hairer_etal_2011} for the numerical simulations.
We adopt $ k_2 = 0.5 $ and $R_\k = 0.25 \, R_\mathrm{Jup}$ for all planets, and $ M = 0.4 \, M_\odot $ (Table\,1).
Typical dissipation times for gaseous planets are $\Delta t_\k \sim 10 $ to $100$\,s, corresponding to dissipation  factors $Q_\k \sim 10^4$ to $10^3$, respectively ($Q_\k^{-1} \approx n_\k \Delta t_\k$).
However, computations with such low $\Delta t_\k$ values (or high $Q_\k$), become prohibitive on account of long evolution times.
Thus, in order to speed up the simulations, in this paper we have considered artificially high values for the tidal dissipation, about one thousand times the expected values ($\Delta t_\k = 10^5 $\,s or $Q_\k \sim 1$).
Time scales are inversely proportional to $\Delta t_\k$ (Eq.\ref{eq.tides}), so 1~Myr of evolution roughly corresponds to 1~Gyr with $\Delta t_\k = 100$\,s (or 10~Gyr with $\Delta t_\k = 10$\,s).

In Figure~\ref{Fdyn7}\,({\it left}) we plot the evolution of the orbital period ratio of the two inner planets together with their orbital eccentricities.
We observe that, although the system remains stable, the eccentricities are progressively damped, while the present period ratio increases towards the 3:1 mean motion resonance, because of the inward migration of the semi-major axes.
Around 0.35\,Myr the system crosses the 3:1 resonance, but capture cannot occur because we have a divergent migration \citep[e.g.][]{Henrard_Lamaitre_1983}.
With a more realistic tidal dissipation ($\Delta t_\k = 10^2 $\,s), this event is scheduled to occur in less than 1~Gyr, so we may wonder why is the present system still evolving in such a dramatic way.

One possibility is that the system is already fully evolved by tidal effect, and the eccentricities of the two inner planets are overestimated (see next section).
Another possibility is to suppose that planet $c$ is terrestrial, since its minimum mass is $6.8 \, M_\oplus$ (Table\,\ref{table:fit}).
Terrestrial planets usually dissipate much more energy than gaseous planets, with typical values $Q_\k \sim 10^1 - 10^2$ \citep[e.g.][]{Goldreich:1966}.
Thus, adopting $\Delta t_c = 5 \times 10^7 $\,s (that is, dissipation for planet $c$ becomes 500 times larger than for the gaseous planets) we repeated the previous simulation, keeping all the other parameters equal, except the initial semi-major axis of this planet: $a_c = 060679$~AU.
In Figure~\ref{Fdyn7}\,({\it middle}) we observe that in this case the orbital period ratio of the two inner planets decreases.
Therefore, the system may have crossed the 3:1 resonance in the past, but evolved to the present situation.
We adopted $a_c$ above the value in the nominal solution (Table\,\ref{table:fit}), so we can see the resonance crossing from above. If we use the nominal value, the orbital period ratio behavior is the same, but decreases to values below the initial 2.97 ratio.

Both the size of the planet and the dissipation rates ($\Delta t$)
are poorly constrained. More generally, the evolution would be longer
for a smaller planet and lower dissipation rates (Eq.~\ref{eq.tides}). For an Earth
composition, planet $c$ minimum mass converts to a radius of roughly 1.7
Rearth \citep{Valencia:2007} and, for the same $\Delta t_c$, the evolution would
take 10 Gyr instead of 1 Gyr. Even for smaller planetary sizes, that
scenario would remain possible if $\Delta t_c$ assumes higher values.

\subsection{Dissipation constraints}

\begin{table}
\caption{Orbital parameters for the planets orbiting GJ\,163, obtained
with a tidal constraint for the proper modes $u_1$ and $u_2$.  \label{tidestab} }
\begin{center}
\begin{tabular}{l l r r r} \hline \hline
\noalign{\smallskip}
{\bf Param.}  & {\bf [unit]} & {\bf b \quad} & {\bf c  \quad} & {\bf d  \quad} \\ \hline 
\noalign{\smallskip}
$\gamma$         & [km/s]             & \multicolumn{3}{c}{58.597}  \\ 
\noalign{\smallskip}
$P$          & [day]                & $ 8.633 $ & $ 25.645 $ & $ 600.895$ \\ 
$\lambda$    & [deg]                & $ 18.252 $ & $ 23.653 $ & $   141.887 $ \\ 
$e$          &                      & $  0.0106 $ & $  0.0094 $ & $   0.3990 $ \\ 
$\omega$     & [deg]                & $ 74.73 $ & $ 235.447$ & $  126.915 $ \\ 
$K$          & [m/s]                & $ 6.121 $ & $  2.901 $ & $   4.711 $ \\ \hline
\noalign{\smallskip}
$\mathrm{m} \sin i$ & [M$_\oplus$]           & $10.661 $    & $7.263 $  & $22.072 $ \\
$a$          & [AU]                 & $0.06069 $           & $ 0.12540 $ 	  & $ 1.02689 $ \\ \hline
\noalign{\smallskip}
$T_\mathrm{epoch}$         & [JD]             & \multicolumn{3}{c}{2\,452\,942.80 (fixed)}  \\ 
$R$     &               &  \multicolumn{3}{c}{50 (fixed)}   \\ 
$u_1$     &               & \multicolumn{3}{c}{$ 0.0275 $}   \\ 
$u_2$     &               & \multicolumn{3}{c}{$ 0.1180 $}   \\ 
$\sqrt{\chi^2}$     &               & \multicolumn{3}{c}{1.52}   \\  
\noalign{\smallskip}
\hline \hline
\end{tabular}
\end{center}
\end{table}

In previous section we saw that the present orbits of the two inner planets in the GJ\,163 are still evolving by tidal effect.
Unless the system started with a much higher value for the eccentricities, and depending on its age, the present eccentricities should have already been damped to lower values.
In addition, dissipation within a primordial disk should have also contributed to circularize the initial orbits \citep[e.g.][]{Papaloizou_2011}.
Thus, it is likely that the eccentricities given by the best fitted solution (Table\,\ref{table:fit}) are overestimated, as it is usual when we use insufficient or inaccurate data \citep[e.g.][]{Pont_etal_2011}.

One can perform a fit fixing both eccentricities $e_b$ and $e_c$ at zero.
This procedure has been done in many previous works, but as explained in \citet{Lovis_etal_2011}, it is not a good approach.
Indeed, if we do so, in the case of the GJ\,163 system, the subsequent evolution of the eccentricities shows a decoupled system ($\Delta \varpi$ is in circulation), where the eccentricities are mainly driven by angular momentum exchanges with the outer planet, and show some irregular variations. 

Over long times, the variations of the planetary eccentricities are usually well described by 
the secular equations (Eqs.\ref{eq.lape}, Figs.\,\ref{Fdyn1},\,\ref{Fdyn2}).
The best procedure to perform a fit to the observational data that takes into account the eccentricity damping constraint is then to make use of these equations.
As for the Laplace-Lagrange linear system (Eqs.\ref{eq.lape}), we can linearize and average the tidal contribution from expression (\ref{eq.tides}) to the eccentricity, and we obtain for each planet $\k$ an additional contribution \citep{Correia_etal_2011}:

\begin{equation}
\dot z_\k = - \gamma_\k \, z_\k \ ,  \quad \gamma_\k = \Delta t_\k \frac{21 k_2 G M^2 R_\k^5}{2 \, \mathrm{m}_\k \,a_\k^8} \ . \label{eq.mar}
\end{equation}
Instead of directly damping the eccentricity, from previous expression it can be shown that tidal effects damp the proper modes $u_\l$ as \citep{Laskar_etal_2012}:
\begin{equation}
u_\l \approx \mathrm{e}^{- \tilde \gamma_\l t} \mathrm{e}^{\mathrm{i} \,(g_\l t + \phi_\l)} \ ,
\quad \tilde \gamma_\l = [\SS^{-1} \mathrm{diag(\gamma_b,\gamma_c,\gamma_d)} \SS]_{\l \l} \ .
\end{equation}

For the present GJ\,163 system, only $\gamma_b $ is relevant.
However, since the inner system is strongly coupled, both proper modes $u_1$ and $u_2$ are damped with $\tilde \gamma_1, \tilde \gamma_2 \sim \gamma_b \approx 10^{-10} $\,yr$^{-1}$ (with $\Delta t_b = 100$\,s), which is compatible with the age of the system.
Dissipation in a primordial disk can add some extra contribution to $\gamma_b$, so we expect proper modes  $u_1$ and $u_2$ to be considerably damped today.
The initial conditions for the GJ\,163 planetary system should then take into account this extra information, similarly to what has been done for the HD\,10180 system \citep{Lovis_etal_2011}.
We have thus chosen to  modify our fitting procedure in order to
include a constraint for the tidal damping of the proper modes $u_1$ and $u_2$, using the additional constraint
\begin{equation}
u_\l = \sum_j \SS_{\l j}^{-1} z_j \approx 0  \ .
\end{equation}
For that purpose, we added to the $ \chi^2 $ minimization, an additional term,
corresponding to these proper modes:
\begin{equation}
\chi^2_R = R \left( u_1^2 + u_2^2 \right) \ ,
\label{chiprop}
\end{equation}
where $R$ is a positive constant, that is chosen arbitrarily in order to obtain
a small value for $ u_1 $ and $ u_2 $ simultaneously.
Using $ R = 50 $ we get $ u_1 \sim 0.03 $ and $ u_2 \sim 0.12 $ and obtain a final $\sqrt{\chi^2} = 1.52$, which is 
nearly identical to the results obtained without this additional constraint 
($R = 0$, $\sqrt{\chi^2} = 1.43$).

The best fit solution obtained by this method is listed in Table\,\ref{tidestab}. 
We believe that this solution is a more realistic representation of the true system than the nominal solution (Table\,\ref{table:fit}).
Indeed, with this constraint, eccentricity variations of the two
innermost planets are regular and slowly damped, while the variations 
in the orbital periods' ratio is almost imperceptible (Fig.\ref{Fdyn7},\,{\it right}). 
In addition, the inner system is still coupled, the two pericentre being locked ($\Delta \varpi = \varpi_c-\varpi_b$ oscillates around $180^\circ$, with a maximal amplitude of about  $26^\circ$).

\subsection{Additional constraints}

\begin{figure*}
    \includegraphics[width=\linewidth]{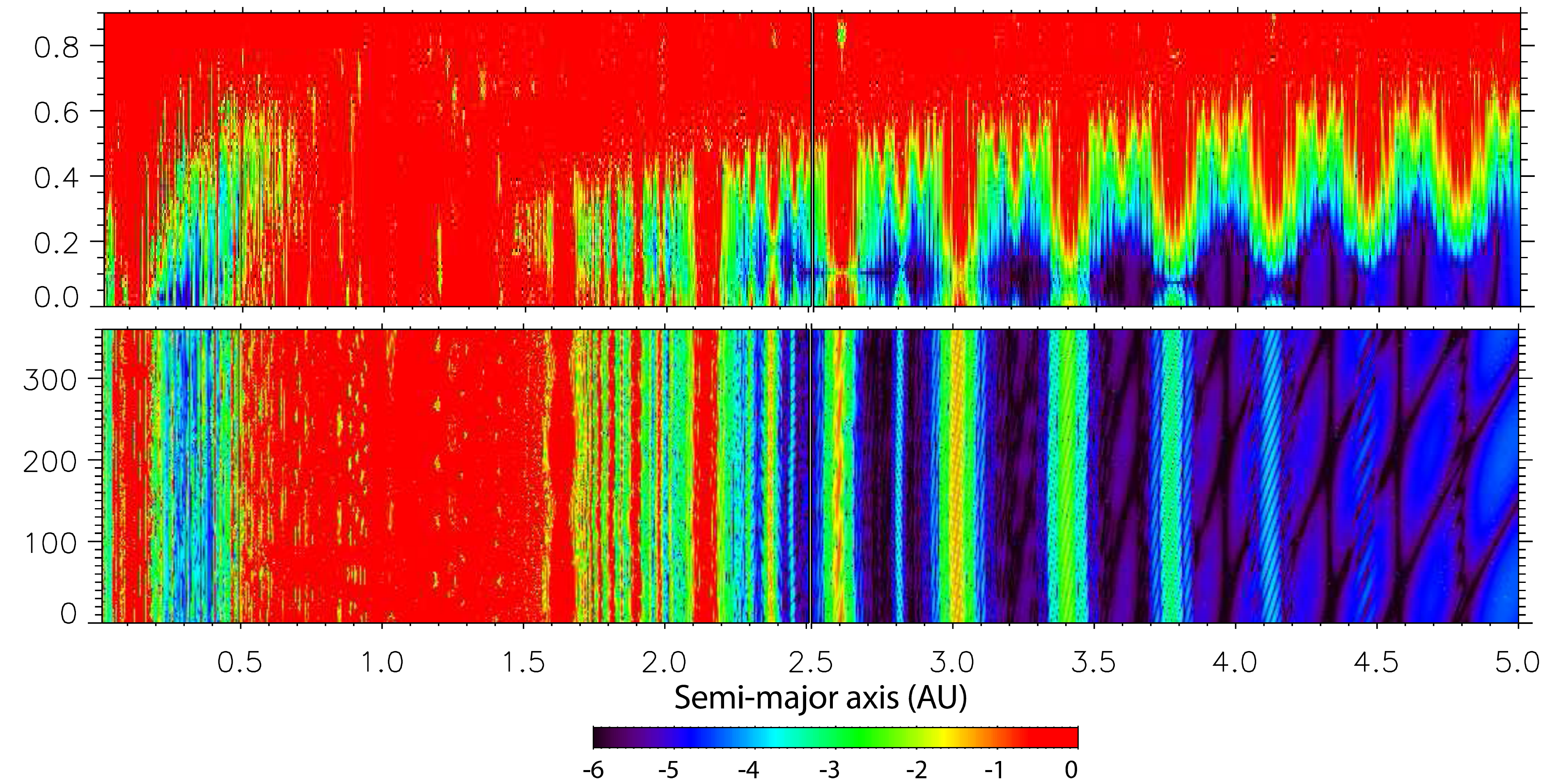} 
  \caption{Possible location of an additional fourth planet in the GJ\,163 system.
  The stability of an Earth-size planet ($K=0.2$~m/s) is analyzed, for various semi-major axis versus 
  eccentricity ({\it top}), or mean anomaly ({\it bottom}). All the angles of the putative planet are set to $0^\circ$ (except the mean anomaly in the bottom panel), and in the bottom panel, its eccentricity to 0.
  The stable zones  where additional planets can be found are the dark blue regions.
    \label{Fdyn8}}   
\end{figure*}

We can assume that the dynamics of the three known planets is not disturbed much by the presence of an additional small-mass planet close-by.
We can thus test the possibility of an additional fourth planet in the system
by varying the semi-major axis, the eccentricity, and the longitude of the
pericentre over a wide range, and performing a stability analysis as in Figure~\ref{Fdyn3}. 
The test was completed for a fixed value  $K=0.2$~m/s, corresponding to
an Earth-mass object at approximately 1~AU, whose radial-velocity amplitude is at the edge of detection (Fig. \ref{Fdyn8}).

From the analysis of the stable areas in Figure~\ref{Fdyn8}, one can see that additional planets are possible beyond 2.5~AU (well outside the outer planet's apocentre), which corresponds to orbital periods longer than 6~yr.
Because the eccentricity of outer planet is high, there are some high-order mean motion resonances that destabilize several zones up to 4~AU.
In addition, the same kind of resonances disturb the inner region between planet $c$ and the pericentre of planet $d$ (Fig.\,\ref{Fdyn6}), although some stability appears to be possible in the range $0.3 < a < 0.5$~AU.
Stability can also be achieved for planets extremely close to the star, with orbital periods shorter than 8~day.

We can also try to find constraints on the maximal masses of the current
three-planet system if we assume co-planarity of the orbits.
Indeed, up to now we have been assuming that the inclination of the system to
the line-of-sight is $90^\circ$, which gives minimum values for the planetary
masses (Table\,\ref{table:fit}).

By decreasing the inclination of the orbital plane of the system, we increase
the mass' values of all planets and repeated a stability analysis of the orbits,
like in Figure~\ref{Fdyn3}.
As we decrease the inclination, the stable ``dark-blue'' areas become narrower,
to a point that the minimum $\chi^2$ of the best fit solution lies outside the
stable zones.
At that point, we conclude that the system cannot be stable anymore.
It is not straightforward to find a transition inclination between the two
regimes, but we can infer from our plots that stability of the whole system is
still possible for an inclination of $30^\circ$, but becomes impossible for an
inclination of of $5^\circ$ or $10^\circ$. 
Therefore, we conclude that the maximum masses of the planets may be
most probably computed for an inclination around $20^\circ$, corresponding to a scaling factor of about 3 for the possible masses.

Even when adopting an inclination of $20^\circ$, the two inner planets lie outside the 3:1 mean motion resonance, more or less at the same place as for $90^\circ$ (Fig.\,\ref{Fdyn3}).
The reason why the system becomes unstable for lower inclination values is because the mass of the outer planet $d$ grows to a point such that high order mean motion resonances between planets  $d$ and $b$ and/or $c$ destroy the whole system.
In particular, the 3:1 resonant island also disappears completely for low inclination values.

\section{\label{sect:hab}Gl163c in the Habitable Zone ?}

With a separation of 0.1254 AU, Gl\,163c receives $\sim$1.34 times more
energy from its star than Earth from the Sun. Considering the case where all the planetary surface re-radiates the absorbed flux
 (e.g. $\beta$ factor of \citet{Kaltenegger:2011b} equal to 1), the equilibrium temperature of
Gl~163c is :
\[
T_{eq} = (302\pm10)(1-A)^{1/4} ~~~{[\rm K]}
\]
Scaled to our Solar System, its
illumination is equivalent to that of a planet located midway between Venus and Earth. 

To be located in the Habitable Zone (HZ), and thus potentially harbor liquid water, the equilibrium temperature of a planet with an atmosphere as dense as Earth
should be between 175K and 270K \citep[see ][for a complete discussion]{Selsis:2007}. In the case of
Gl\,163c such condition is fulfilled for large range of Bond albedos ($A=0.34-0.89$), but not for an albedo similar to that of Earth. The albedo of Earth is equal to 0.3 in the optical and is as low as 0.2 in the near-IR, where early M dwarfs radiates most of their energy. With such values, Gl\,163c would lie outside the HZ. An albedo greater than 0.34 is however possible if 40-50\% of the atmosphere is covered by clouds \citep[see for example Fig.~1 of][]{Kaltenegger:2011b}. The precise location of GJ\,163c with respect to the habitable zone may further depend on additional heating such as tidal \citep{Barnes:2012a} or radiogenic \citep{Ehrenreich:2006a} heatings, and more detailed studies are thus welcome.

Two other conditions are needed for a planet in the HZ to be really
habitable (e.g. host liquid water on its surface). Firstly the planet
should not have accreted a massive H-He envelope, otherwise the surface
pressure would be too strong and may lead to a runaway greenhouse effect. On the 3-10M$_{\odot}$ range planets can
have very different structures for a given mass, and it is impossible
to know, without a radius measurement, whether GJ\,163c is embedded in a
massive H-He envelope or not. Secondly the planet should contain water among
the component of its atmosphere.

On the case of planets inside HZ
around M dwarfs numerous discussions exist about two particularity and
their effect on the habitability : their location inside the tidal lock
radius of their star and the high activity level of M dwarfs. In
\citet{Delfosse:2012} we summarized the results of recent works on this
domain. The main conclusion is that tidal effect and atmospheric
erosion from the neighborhood of active stars does not "preclude the
habitability of terrestrial planets orbiting around cool stars
\citep{Barnes:2010}". In particular, the thick atmosphere that may enshroud a planet
of $\sim$7 Earth-mass seem stable even around very active M dwarfs
\citep{Tian:2009}.

\section{\label{sect:concl}Conclusion}
We have presented the analysis of 150 HARPS RVs of the nearby M dwarf GJ\,163 and demonstrated it encodes at least 3 signals compatible with the gravitational pull of orbiting planets and identified 2 additional signals that need further observations before counting them as additional planets. Signals b and d have periodicities that seem incompatible with the possible rotational periods of the star. Signals b and c are also recovered when the data set is divided in observational season, lending credence that at least 3 planets orbit around GJ\,163. We derived their orbital periods ($\sim$8.6, 25.6 and 604 days) and their minimum masses ($\sim$10.6, 6.8 and, 29 ${\rm M_\oplus}$), which correspond to a hot, a temperate and a cold planet in the super-Earth/Neptune mass regime. The super-Earth GJ\,163c may retain further attention for its potential habitability. It receives about 30\% more energy than Earth in our Solar System and could qualify as a habitable-zone planet for a wide range of albedo values ($175 \le T_{eq} \le 270$~K, for $0.34 \le A \le 0.89$).

We also performed a detailed dynamical analysis of the system to show that, despite a period ratio $P_c/P_b=2.97$, planets b and c do not participate in a 3:1 resonance. The system is found stable over a time comparable to the age of the system and, as far as the orbital parameters of the first three planets remain unchanged, it also appears complete down to Earth-mass planets for a wide range of separations (0.1 $\lesssim a \lesssim $ 2.2 AU).

GJ\,163 is a singular system both for its potentially habitable planets, GJ\,163c, and, for its particular hierarchal structure and dynamical history. And therefore, before its atmosphere can be characterized and search for biomarkers with future observatories, it is already a unique system to theoretically connect the potential habitability of a planet with the dynamical history of a planetary system.

\begin{acknowledgements}
Our first thanks go to the ESO La Silla staff which we are grateful for its continuous support. We wish to thank the anonymous referee for thoughtful comments and suggestions. We also acknowledge support by PNP-CNRS, CS of Paris Observatory,
PICS05998 France-Portugal program, Funda\c{c}\~ao para a Ci\^encia e a Tecnologia (FCT) through program Ci\^encia\,2007 funded by FCT/MCTES (Portugal) and POPH/FSE (EC) and, by the European Research Council/European Community under the FP7 through a Starting Grant (grants PTDC/CTE-AST/098528/2008, PTDC/CTE-AST/098604/2008, PEst-C/CTM/LA0025/2011 and, SFRH/BD/60688/2009), grant agreement 239953. MG is FNRS Research Associate.
\end{acknowledgements}

\bibliographystyle{aa}
\bibliography{mybib2}

\begin{table*}[t!]
\caption{\label{TabOrb2}Fitted orbital solution for the GJ\,163 
  planetary system: 5 Keplerians. Keplerian signals labeled $b$, $c$ and $d$ are interpreted as due to orbiting planets whereas those labeled $(e)$ and $(f)$ require more data for robust interpretations (and are not considered as planet detections at this stage; see Sect.~\ref{sect:data} \&~\ref{sect:activity}).
  }

\begin{tabular}{l l l c c c c c}
  \hline\hline
  \multicolumn{2}{l}{\bf Parameter} &\hspace*{2mm} 
  & \bf GJ\,163\,b  & \bf GJ\,163\,(e)  & \bf GJ\,163\,c & \bf GJ\,163\,(f) & \bf GJ\,163\,d \\
  \hline
  $P$ & [days] & & 8.631 $\pm$ 0.001 & 19.46 $\pm$ 0.02 & 25.60 $\pm$ 0.02 & 108.4 $\pm$ 0.5 & 603$\pm$12\\
  $T$ & [JD-2400000] & & 55042.553 $\pm$ 0.01 &  55040.5 $\pm$ 1.4 & 55057.7 $\pm$ 4.8 & 55037.6 $\pm$ 5.6 & 55275$\pm$24 \\
  $e$ & & & 0.11$\pm$0.04 & 0.32$\pm$0.17 & 0.08 $\pm$   0.08 & 0.41$\pm$   0.15  & 0.41$\pm$0.07\\
  $\omega$ & [deg] & & 97$\pm$22  & 125 $\pm$ 31& -145 $\pm$ 68 & 141 $\pm$  23 & 88$\pm$17\\
  $K$ & [m s$^{-1}$] & & 6.22 $\pm$ 0.26 & 1.69 $\pm$ 0.32 & 3.07 $\pm$ 0.27 & 1.94 $\pm$ 0.38 & 3.82$\pm$0.38\\
  $V$ & [km s$^{-1}$] & & \multicolumn{4}{c}{58.5965 $\pm$ 0.0008}  \\
  $m_2 \sin{i}$ & [$M_{\oplus}$] & & 11 & 3.7 & 7.7 & 7.2 & 25 \\
  $a$ & [AU] & & 0.061 & 0.10 & 0.13 & 0.33 &1.0\\
  \hline
  $N_{\mathrm{meas}}$ & & & \multicolumn{5}{c}{153} \\
  {\it Span} & [days] & & \multicolumn{5}{c}{3068} \\
  $\sigma$ (O-C) & [ms$^{-1}$] & & \multicolumn{5}{c}{2.02} \\
  $\chi^2_{\rm red}$ & & & \multicolumn{5}{c}{1.21} \\
  \hline
\end{tabular}

\end{table*}

\longtab{6}{\begin{longtable}{lllllll}
\caption{\label{tab:rv} Radial velocity time series of GJ\,163, given in the Solar System barycentric reference frame (the secular acceleration due to GJ\,163 proper motion is not removed), together with measurements of the full width half maximum (FWHM) and bisector span (BIS) of the cross-correlation function, as well as \ion{Ca}{ii} H+K and H$\alpha$ indices.}\\
\hline\hline
BJD-2400000.0 & RV [km/s] & $\sigma_{\rm RV}$[km/s] & FWHM [km/s] & BIS [km/s] & \ion{Ca}{ii} H+K & H$\alpha$ \\ \hline \endfirsthead
\caption{continued.}\\
\hline\hline
BJD-2400000.0 & RV [km/s] & $\sigma_{\rm RV}$ [km/s] & FWHM [km/s] & BIS [km/s] & \ion{Ca}{ii} H+K & H$\alpha$\\ \hline \endhead \hline \endfoot
52942.803917	&58.599730	&0.002140	&3.532460	&-0.010860	&0.474612	&1.083375\\
52991.725755	&58.591194	&0.002540	&3.532960	&-0.018580	&0.316633	&1.072997\\
52998.645887	&58.585455	&0.002440	&3.528400	&-0.006190	&0.398500	&1.076006\\
52999.691817	&58.585924	&0.002770	&3.554450	&-0.013840	&0.459171	&1.077594\\
53000.594071	&58.593522	&0.002050	&3.537920	&-0.011000	&0.260918	&1.081138\\
53000.722076	&58.591602	&0.002370	&3.545950	&-0.011240	&0.374872	&1.070918\\
53002.617008	&58.595600	&0.002800	&3.538410	&-0.003550	&0.270616	&1.075801\\
53007.604241	&58.585183	&0.001390	&3.536030	&-0.012540	&0.394607	&1.078777\\
53788.544601	&58.608563	&0.001620	&3.541870	&-0.005620	&0.230016	&1.079363\\
53980.880274	&58.598335	&0.001840	&3.546480	&-0.013860	&0.360284	&1.073684\\
53989.894496	&58.600972	&0.002610	&3.549480	&-0.008850	&0.526464	&1.075047\\
54384.865244	&58.601831	&0.002750	&3.552940	&-0.004280	&0.801770	&1.078881\\
54430.698362	&58.595520	&0.001930	&3.545380	&-0.010040	&0.458242	&1.074572\\
54437.718735	&58.594390	&0.001790	&3.546480	&-0.008590	&0.169284	&1.079625\\
54438.653391	&58.591809	&0.001950	&3.550160	&-0.007970	&0.369221	&1.078956\\
54478.650697	&58.606175	&0.001770	&3.542980	&-0.011420	&0.440735	&1.072577\\
54487.573093	&58.601493	&0.001840	&3.546970	&-0.014810	&0.465068	&1.080293\\
54520.609369	&58.596919	&0.001980	&3.543680	&-0.012410	&0.551981	&1.074354\\
54719.896111	&58.588711	&0.002200	&3.556560	&-0.009990	&0.531249	&1.077615\\
54731.888605	&58.589755	&0.001410	&3.543010	&-0.010460	&0.326511	&1.089013\\
54733.816378	&58.586632	&0.001490	&3.547850	&-0.003550	&0.353112	&1.073906\\
54751.810852	&58.581438	&0.002900	&3.555870	&-0.008810	&0.563352	&1.072038\\
54812.616009	&58.587256	&0.001940	&3.558610	&-0.009410	&0.238095	&1.070079\\
54825.566352	&58.589809	&0.001380	&3.544240	&-0.010290	&0.217615	&1.084402\\
54826.578927	&58.586028	&0.001520	&3.538940	&-0.009790	&0.380201	&1.074107\\
54827.608005	&58.580606	&0.001480	&3.543070	&-0.007420	&0.294079	&1.080112\\
54828.620174	&58.581445	&0.001800	&3.545750	&-0.009490	&0.430507	&1.074123\\
54829.586094	&58.583984	&0.001930	&3.547440	&-0.006120	&0.234094	&1.078419\\
54830.613066	&58.588022	&0.001900	&3.552760	&-0.010410	&0.304014	&1.073584\\
54831.607503	&58.592091	&0.001560	&3.552750	&-0.010580	&0.362226	&1.074913\\
54832.618387	&58.597760	&0.002350	&3.546830	&-0.018280	&0.263249	&1.077878\\
54833.609328	&58.593568	&0.002070	&3.554360	&-0.011130	&0.613232	&1.071895\\
54834.652456	&58.597337	&0.001510	&3.543960	&-0.011310	&0.412504	&1.084964\\
54840.627291	&58.594869	&0.001680	&3.551160	&-0.012560	&0.379793	&1.076521\\
54848.543610	&58.590978	&0.001980	&3.555530	&-0.012560	&0.349984	&1.076927\\
54849.552882	&58.592307	&0.001360	&3.545530	&-0.012440	&0.391225	&1.076725\\
54850.566102	&58.593265	&0.001790	&3.550930	&-0.008870	&0.400276	&1.081782\\
54851.610882	&58.592164	&0.001460	&3.541880	&-0.012200	&0.409678	&1.077396\\
54852.619391	&58.586663	&0.001600	&3.544490	&-0.011980	&0.294328	&1.081587\\
54854.600264	&58.586580	&0.001740	&3.552600	&-0.016610	&0.383535	&1.078345\\
54871.562478	&58.587167	&0.001670	&3.557320	&-0.004500	&0.450362	&1.080230\\
54878.555254	&58.592398	&0.001810	&3.548870	&-0.009940	&0.333256	&1.083030\\
54879.577238	&58.586426	&0.001860	&3.539120	&-0.001320	&0.422477	&1.075878\\
54880.555427	&58.586315	&0.001570	&3.542120	&-0.012100	&0.398863	&1.072781\\
54881.611126	&58.589184	&0.002090	&3.542500	&-0.010160	&0.263213	&1.063036\\
54882.557176	&58.595332	&0.001820	&3.539640	&-0.010240	&0.273284	&1.081501\\
54883.549162	&58.600011	&0.001740	&3.539350	&-0.011080	&0.317384	&1.074110\\
54884.576564	&58.602030	&0.002040	&3.540710	&-0.011110	&0.084204	&1.076239\\
54885.528112	&58.603828	&0.001590	&3.541850	&-0.012140	&0.231690	&1.075653\\
54886.533363	&58.595717	&0.001970	&3.532180	&-0.016230	&0.464079	&1.071355\\
54914.550445	&58.591029	&0.002690	&3.549890	&-0.009630	&0.471671	&1.078348\\
54916.508564	&58.597247	&0.001740	&3.544650	&-0.013670	&0.441040	&1.077088\\
54918.502657	&58.603854	&0.001680	&3.547320	&-0.011830	&0.307162	&1.075228\\
54920.528430	&58.599961	&0.001430	&3.540750	&-0.013140	&0.474758	&1.079531\\
54939.469067	&58.591066	&0.001710	&3.549860	&-0.011120	&0.390615	&1.076822\\
54940.478788	&58.586375	&0.002220	&3.543790	&-0.008430	&0.399146	&1.074327\\
54941.476414	&58.590163	&0.001430	&3.544420	&-0.007030	&0.461203	&1.073636\\
54946.478962	&58.597626	&0.001660	&3.536830	&-0.013740	&0.318449	&1.077012\\
55041.918339	&58.600488	&0.002190	&3.550690	&-0.006260	&0.480391	&1.075341\\
55042.886725	&58.592837	&0.001680	&3.543100	&-0.014170	&0.360061	&1.075193\\
55044.896707	&58.591394	&0.002700	&3.555330	&-0.008500	&0.579905	&1.079231\\
55045.911345	&58.596453	&0.002610	&3.555330	&-0.005500	&0.285818	&1.076652\\
55046.894559	&58.592392	&0.003600	&3.542700	&-0.008000	&0.415718	&1.074274\\
55047.902460	&58.596590	&0.002630	&3.552630	&-0.008210	&0.565590	&1.077283\\
55048.885477	&58.595589	&0.002630	&3.547930	&-0.002660	&0.038106	&1.065723\\
55049.886410	&58.601887	&0.001960	&3.554390	&-0.007880	&0.325910	&1.082535\\
55052.879997	&58.586763	&0.001660	&3.541390	&-0.012790	&0.310553	&1.080782\\
55053.913373	&58.589372	&0.002640	&3.553330	&-0.003510	&0.398575	&1.085811\\
55054.912477	&58.591741	&0.002200	&3.544320	&-0.007530	&0.463345	&1.076842\\
55063.917531	&58.595689	&0.001890	&3.555740	&-0.015280	&0.276842	&1.088665\\
55065.915353	&58.602476	&0.001300	&3.541460	&-0.009280	&0.420095	&1.078647\\
55067.918464	&58.601273	&0.001310	&3.539850	&-0.015490	&0.260674	&1.076205\\
55071.889831	&58.597578	&0.001880	&3.558080	&-0.008980	&0.277670	&1.074310\\
55072.894105	&58.599877	&0.001660	&3.548190	&-0.007030	&0.323137	&1.080621\\
55073.907719	&58.605925	&0.001630	&3.555040	&-0.012500	&0.402864	&1.081054\\
55074.919642	&58.601454	&0.001880	&3.558770	&-0.008840	&0.252327	&1.077207\\
55075.895014	&58.602353	&0.002300	&3.556800	&-0.010730	&0.201046	&1.078968\\
55077.892991	&58.593700	&0.002390	&3.557860	&-0.008840	&0.283597	&1.073395\\
55091.903973	&58.604361	&0.002050	&3.554380	&-0.013920	&0.496218	&1.075653\\
55095.879622	&58.594076	&0.001420	&3.536210	&-0.011500	&0.296341	&1.071715\\
55097.887687	&58.595463	&0.001950	&3.553440	&-0.004090	&0.463611	&1.075522\\
55100.902049	&58.601319	&0.001620	&3.545710	&-0.017500	&0.354803	&1.071907\\
55101.880257	&58.598838	&0.002440	&3.550360	&-0.004930	&0.287275	&1.074712\\
55105.790168	&58.591182	&0.002230	&3.554200	&-0.015500	&0.288853	&1.073228\\
55106.871732	&58.594661	&0.002080	&3.557470	&-0.009200	&0.296981	&1.070556\\
55113.820733	&58.598102	&0.001500	&3.548620	&-0.018730	&0.342577	&1.074824\\
55116.822247	&58.605618	&0.001920	&3.556280	&-0.003390	&0.334569	&1.077636\\
55121.802467	&58.592791	&0.002070	&3.543880	&-0.014070	&0.556809	&1.072763\\
55122.787181	&58.591129	&0.002600	&3.553800	&-0.011860	&0.342499	&1.082588\\
55123.736963	&58.594078	&0.002060	&3.548640	&-0.007990	&0.403123	&1.075784\\
55124.852544	&58.599667	&0.001570	&3.538870	&-0.010590	&0.380541	&1.078000\\
55126.822581	&58.601534	&0.002100	&3.538750	&-0.015420	&0.410731	&1.079145\\
55128.842315	&58.594721	&0.001550	&3.543490	&-0.010950	&0.392741	&1.082422\\
55129.825393	&58.591950	&0.001470	&3.541970	&-0.008200	&0.407386	&1.074676\\
55132.859403	&58.597676	&0.002430	&3.543970	&-0.014570	&0.582579	&1.076822\\
55133.777872	&58.600445	&0.002350	&3.550690	&-0.013620	&0.458441	&1.077885\\
55134.841804	&58.602523	&0.001680	&3.548870	&-0.008590	&0.210445	&1.077906\\
55135.648814	&58.602442	&0.001750	&3.548870	&-0.011990	&0.440236	&1.081918\\
55137.656304	&58.592419	&0.001930	&3.546950	&-0.010010	&0.142718	&1.075953\\
55138.746506	&58.587788	&0.001980	&3.553610	&-0.007680	&0.352444	&1.072225\\
55139.661813	&58.589477	&0.002440	&3.550780	&-0.012500	&0.299936	&1.069435\\
55140.748449	&58.594495	&0.001420	&3.544860	&-0.010880	&0.418191	&1.070070\\
55141.686972	&58.595514	&0.001750	&3.551000	&-0.013440	&0.234147	&1.075958\\
55142.699430	&58.601173	&0.002820	&3.543630	&-0.014510	&0.478706	&1.073814\\
55160.662923	&58.600719	&0.001770	&3.545430	&-0.015580	&0.443681	&1.078016\\
55161.646551	&58.599377	&0.002650	&3.539930	&0.001980	&0.468896	&1.077736\\
55162.644184	&58.601206	&0.001930	&3.546740	&-0.007850	&0.222262	&1.070983\\
55163.651074	&58.597565	&0.001900	&3.543830	&-0.009980	&0.316073	&1.075364\\
55164.616403	&58.590343	&0.001580	&3.545510	&-0.013860	&0.383956	&1.073808\\
55165.580065	&58.593162	&0.001620	&3.546550	&-0.016210	&0.358046	&1.078572\\
55166.608077	&58.597381	&0.001440	&3.545010	&-0.012820	&0.385386	&1.072063\\
55167.603207	&58.603409	&0.001410	&3.544360	&-0.013290	&0.395243	&1.070626\\
55168.570652	&58.603138	&0.001610	&3.547720	&-0.010480	&0.392509	&1.072981\\
55169.596567	&58.609967	&0.001550	&3.544590	&-0.010300	&0.442084	&1.076053\\
55217.675575	&58.590562	&0.001550	&3.547150	&-0.007040	&0.363191	&1.077256\\
55218.667684	&58.597251	&0.001620	&3.548580	&-0.012870	&0.420014	&1.074939\\
55230.568213	&58.611425	&0.002170	&3.545790	&-0.010580	&0.394349	&1.076228\\
55233.523965	&58.588131	&0.002210	&3.544910	&-0.010080	&0.385514	&1.073791\\
55236.517841	&58.591807	&0.001900	&3.542950	&-0.009810	&0.211261	&1.073849\\
55260.527223	&58.581074	&0.001780	&3.546110	&-0.013550	&0.411877	&1.082340\\
55261.560450	&58.589773	&0.002040	&3.557030	&-0.010440	&0.372070	&1.070446\\
55423.831054	&58.595135	&0.003040	&3.548410	&-0.012580	&-0.265765	&1.075284\\
55437.824428	&58.595176	&0.001560	&3.544680	&-0.012990	&0.378742	&1.076729\\
55453.906581	&58.606984	&0.001810	&3.554740	&-0.008550	&0.302701	&1.085761\\
55454.881206	&58.601113	&0.002100	&3.552760	&-0.017370	&0.367568	&1.078111\\
55456.881788	&58.594540	&0.001740	&3.546090	&-0.009510	&0.426824	&1.073723\\
55458.898156	&58.589288	&0.002100	&3.543120	&-0.014250	&0.237104	&1.083107\\
55483.779915	&58.581704	&0.003220	&3.562560	&-0.008910	&0.472817	&1.074149\\
55484.760395	&58.582863	&0.002180	&3.548450	&-0.021400	&0.326292	&1.074936\\
55485.779369	&58.587362	&0.001980	&3.549790	&-0.016300	&0.304048	&1.072448\\
55487.780869	&58.589239	&0.001940	&3.546180	&-0.012820	&0.279858	&1.073912\\
55489.705353	&58.592336	&0.001700	&3.538230	&-0.009840	&0.396902	&1.075609\\
55538.588842	&58.593771	&0.002220	&3.541780	&-0.016770	&0.296814	&1.079429\\
55541.614985	&58.600546	&0.001660	&3.548690	&-0.010100	&0.277661	&1.077177\\
55545.681700	&58.588501	&0.001810	&3.545320	&-0.016160	&0.309868	&1.078832\\
55580.622938	&58.594244	&0.001720	&3.547430	&-0.013150	&0.328606	&1.073628\\
55629.504390	&58.595658	&0.002010	&3.545060	&-0.007120	&0.200347	&1.076790\\
55824.896015	&58.595576	&0.002370	&3.551940	&-0.017700	&0.455140	&1.083222\\
55829.829382	&58.592019	&0.001550	&3.545150	&-0.006660	&0.440034	&1.082638\\
55837.818457	&58.595558	&0.001460	&3.542820	&-0.007690	&0.278344	&1.077121\\
55844.837978	&58.596689	&0.001870	&3.557030	&-0.008390	&0.282813	&1.077810\\
55872.669291	&58.589611	&0.002190	&3.555520	&-0.013860	&0.397152	&1.072780\\
55874.752398	&58.593999	&0.002050	&3.561370	&-0.011840	&0.434294	&1.076499\\
55880.757925	&58.589481	&0.001480	&3.545590	&-0.009220	&0.491712	&1.080612\\
55889.659774	&58.586679	&0.001950	&3.549130	&-0.009890	&0.306915	&1.072234\\
55891.661811	&58.591376	&0.001750	&3.543730	&-0.010430	&0.482063	&1.080936\\
55894.673249	&58.598942	&0.001520	&3.545020	&-0.009550	&0.429553	&1.077865\\
55924.612672	&58.583442	&0.001590	&3.552630	&-0.011340	&0.355721	&1.083856\\
55930.642696	&58.592063	&0.001980	&3.547990	&-0.023220	&0.548718	&1.080222\\
55941.536494	&58.588739	&0.001760	&3.552610	&-0.008000	&0.284015	&1.082646\\
55999.492554	&58.593711	&0.001720	&3.546750	&-0.012300	&0.388007	&1.079473\\
56002.486731	&58.583657	&0.002480	&3.557530	&-0.002560	&0.424503	&1.064315\\
56010.500818	&58.586336	&0.001750	&3.540990	&-0.011850	&0.470636	&1.082683\\
\end{longtable}}

\end{document}